\documentclass[journal,twocolumn]{IEEEtran}
\IEEEoverridecommandlockouts

\ifCLASSINFOpdf

\else

\fi

\usepackage{booktabs}
\usepackage{enumitem}

\usepackage{bbm}
\usepackage{bigints}
\usepackage{amsmath}
\usepackage{amssymb}
\usepackage{outlines}
\usepackage{array}
\usepackage{float}
\usepackage{kotex}
\usepackage{graphicx}
\usepackage{bm}
\usepackage{caption,subcaption}
\usepackage[font=footnotesize]{caption}
\usepackage{algorithm}
\usepackage{algpseudocode}
\usepackage{csquotes}
\usepackage{cite}
\usepackage{multicol}
\usepackage{stmaryrd}
\usepackage{tabularx,booktabs}
\usepackage{xcolor}
\newcolumntype{C}{>{\centering\arraybackslash}X} 
\usepackage{lipsum} 
\usepackage{multirow}
\usepackage{caption,subcaption}
\usepackage[
            top=0.75in,
            bottom=1in,
            left=0.625in,
            right=0.625in]{geometry}
\newsavebox{\measurebox}

\begin{document}
%
\title{Exploring the Near and Far-Field Coexistence for RIS-Assisted ISAC Systems: An Adaptive Bandwidth Splitting Approach}
%
%
%

\author{Seonghoon~Yoo,~\IEEEmembership{Graduate Student Member,~IEEE,} Jaemin~Jung,\\Seongah~Jeong,~\IEEEmembership{Senior Member,~IEEE,} Jinkyu~Kang,~\IEEEmembership{Member,~IEEE,}\\ Markku Juntti,~\IEEEmembership{Fellow,~IEEE,} and
Joonhyuk~Kang,~\IEEEmembership{Member,~IEEE}
\thanks{This work was partly supported by Institute of Information \& communications Technology Planning \& Evaluation (IITP) grant funded by the Korea government (MSIT) (No. RS-2024-00444230, Development of Wireless Technology for Integrated Sensing and Communication) and the National Research Foundation of Korea (NRF) grant funded by the Korea government (MSIT) (No. RS-2023-NR077102). \textit{(Corresponding author: Jinkyu Kang and Joonhyuk Kang)}}
\thanks{Seonghoon Yoo and Joonhyuk Kang are with the School of Electrical Engineering, Korea Advanced Institute of Science and Technology, Daejeon 34141, South Korea (e-mail: shyoo902@kaist.ac.kr, jhkang@ee.kaist.ac.kr).}
\thanks{Jaemin Jung and Jinkyu Kang are with the Department of Information and
Communication Engineering, Myongji University, Gyeonggi-do 17058, South
Korea (e-mail: jjm91548971@mju.ac.kr, jkkang@mju.ac.kr).}
\thanks{Seongah Jeong is with the School of Advanced Fusion Studies, Department of Intelligent Semiconductor Engineering, University of Seoul, Seoul 02504,
South Korea (e-mail: seongah@uos.ac.kr).}
\thanks{Markku Juntti is with the University of Oulu, 90570 Oulu, Finland (email:
markku.juntti@oulu.fi).}}

\markboth{IEEE Transactions on Wireless Communications,~Vol.~xx, No.~xx, XXX~2025}%
{Shell \MakeLowercase{\textit{et al.}}: Bare Demo of IEEEtran.cls for IEEE Journals}

\maketitle

\begin{abstract}
Integrated sensing and communication (ISAC) enables joint use of spectrum and hardware resources for radar sensing and data transmission, serving as a key enabler of next-generation wireless networks. However, most existing ISAC studies have been limited to operation within a single frequency band and have not been designed to adapt to diverse wireless propagation environments or user configurations. To address these limitations, this paper investigates a reconfigurable intelligent surface (RIS)-assisted ISAC system employing an adaptive bandwidth-splitting strategy under near-field (NF) and far-field (FF) coexistence. The system comprises a full-duplex access point (AP), an RIS and multiple users, where an ISAC user (IU) is a sensing target as well as communication user in the NF region, while communication-only users (CUs) rely on RIS and experience either NF or FF propagation depending on its placement. The proposed system jointly exploits the traditional sensing-only (SO) and ISAC bands and adopts uplink non-orthogonal multiple access (NOMA) for simultaneous transmission. We formulate a joint optimization problem for the receive beamforming vector, bandwidth-splitting ratio and RIS phase shifts to minimize the Cramér–Rao bound (CRB) under rate and resource constraints. An efficient algorithm is developed based on an alternating optimization (AO) framework combined with the semi-definite relaxation (SDR). Numerical results demonstrate that the proposed approach significantly outperforms conventional schemes that operate solely in either the ISAC or SO band, achieving superior performance across various RIS and user configurations under hybrid NF and FF coexistence scenarios.

\end{abstract}


\IEEEpeerreviewmaketitle

\section{Introduction}

\IEEEPARstart{T}{he} advent of sixth-generation (6G) wireless networks is expected to significantly enhance massive connectivity by leveraging the abundant spectrum in mmWave or sub-THz bands \cite{ISAC1, ISAC2}. In particular, future wireless networks are envisioned to accommodate diverse communication and sensing requirements arising from a wide range of user types and operating environments. Among various approaches to support such heterogeneous users, conventional network-based positioning or sensing methods often experience performance degradation due to multipath effects and interference in diverse propagation environments \cite{xu2024isac_vs_conv}. By contrast, integrated sensing and communication (ISAC) systems provide a unified framework that employs radar sensing techniques, enabling not only accurate localization but also mobility and direction estimation by leveraging the information of echo channels explicitly modeled for sensing targets, which is essential for future wireless systems \cite{ISAC1, ISAC2}. 

Recent studies have intensively investigated reconfigurable intelligent‑surface (RIS)‑aided ISAC systems to achieve further performance enhancements by adjusting the amplitude and phase shift of each element and reconfiguring the wireless propagation environment \cite{Ruizhang_ISAC, Cho2024RIS, Yoo2024ISAC, NF_RIS, near_ris_isac,near_ris_isac_2, NF_ISAC,  Wang2024RIS_nf_ff_ISAC,ris_near_far_coexistence,  Luo2025RIS_ISAC_NEAR_FAR}. In \cite{Ruizhang_ISAC}, an RIS‑enabled ISAC system is presented to achieve the enhanced sensing accuracy for multiple users in the far-field (FF) communications with the planar wave channel model. However, for the upcoming 6G networks with the use of a massive number of antennas and the dramatically increased carrier frequency in millimeter wave (mmWave) and terahertz (THz) bands, the near-field (NF) region cannot be neglected as in the past generation. According to the definition of Rayleigh distance, also called the Fraunhofer distance \cite{Selvan2017Fraunhofer}, in the multiple-input multiple-output (MIMO) scenario with the $0.36$ m antenna aperture at $28$ GHz, the NF range becomes $100$ m. In addition, when the RIS is employed between the access point (AP) and the users, the inter-node distance decreases, and the NF propagation with the spherical wave needs to be considered for the future 6G, which becomes indispensable in RIS-aided ISAC systems. To this end, \cite{NF_RIS} extends the channel modeling framework by introducing a cascaded NF model for RIS-assisted links, and \cite{near_ris_isac} demonstrates enhanced estimation accuracy when the RIS is placed in close proximity to both users and targets. Although NF models impose stringent computational and estimation requirements, they can theoretically yield superior sensing and communication performances than counterparts that only consider FF channels \cite{NF_ISAC}. Moreover, the recent works \cite{ris_near_far_coexistence, NF_FF_coexistence, Xiao2025NF_FFcoexistence} have initiated the study of hybrid NF and FF coexistence channel models, where the NF beam-focusing enhances estimation accuracy for close-range targets, while the FF plane‑wave model is employed beyond the focal range by effectively serving multiple users within a cone-shaped beam. Accordingly, the coexistence of NF and FF communications becomes an essential component for the design of RIS‑aided ISAC systems, which requires revisiting the conventional system design to accommodate both field conditions.

In practical wireless environments, the coexistence of NF and FF regions naturally arises, as users and sensing targets are distributed across a wide range of distances from the AP. Furthermore, the placement or movement of RIS introduces mixed propagation characteristics depending on its relative distance from the users, which cannot be adequately captured by a single-field assumption. For example, in indoor industrial scenarios, e.g., smart manufacturing plants and automated warehouses, RIS can assist manufacturing or monitoring devices located in the FF region of the AP, while AGVs \cite{Indoor_ISAC,IIoT_noma} and logistics devices \cite{automated_guided_vehicles} operating in the NF region can additionally benefit from sensing capabilities. In outdoor environments, such as urban transportation networks with dense traffic and roadside infrastructures, the AP can sense the positions of vehicles or unmanned aerial platforms (UAVs) \cite{Yang2024Vehicle_TWC} in the NF region, and simultaneously support users in the FF region with the aid of RIS deployed into roadside monitoring terminals. These diverse deployment conditions highlight the need for flexible ISAC frameworks capable of adapting to spatial variations and hybrid propagation characteristics.

Within the RIS-assisted ISAC framework, since sensing echoes and communication waveforms are simultaneously received over the same radio frequency band in both NF and FF scenarios, a suitable multiple-access mechanism is crucial to ensure reliable signal decoding. Non-orthogonal multiple access (NOMA) has emerged as a promising solution, as it enables the superposition of multiple signals within a single frequency band and facilitates their separation through successive interference cancellation (SIC) \cite{ Wang2022noma_isac, Xue2025noma_isac2, noma_ris_isac_1, noma_ris_isac_2, noma_ris_isac_3}. The studies in \cite{Wang2022noma_isac, Xue2025noma_isac2} propose NOMA-enabled ISAC frameworks that manage the interference between sensing and communication functionalities, while the authors in \cite{noma_ris_isac_1, noma_ris_isac_2, noma_ris_isac_3} demonstrate that cluster-based NOMA within RIS-aided ISAC systems can effectively enhance spectral efficiency. More recently, \cite{partialNOMA2023Akhtar} and \cite{Semi_ISAC} introduce analytical frameworks that evaluate sensing accuracy under bandwidth splitting between ISAC bands and dedicated sensing or communication bands, thereby providing useful insights into NOMA-assisted ISAC design. However, these works do not explicitly address the problem of optimal bandwidth allocation, which becomes critical in RIS-assisted ISAC systems, where dynamic transitions between NF and FF continuously affect the balance between sensing and communication performance. In such complex propagation environments, adaptively optimizing the bandwidth-splitting ratio is essential to achieve a balanced trade-off between sensing accuracy and communication efficiency under varying user configurations and channel conditions.

\begin{table*}[t]
    \centering
    \caption{Technical Features of Representative Works on RIS-assisted ISAC}
    \label{related works}
    \begin{tabular}{lcccccl}
        \toprule
        \textbf{References}  & \textbf{NOMA} & \textbf{RIS} &
        \textbf{Bandwidth Split} & \textbf{Target Scenario} & \textbf{Channel Model} \\
        \midrule
        
        
        \cite{ISAC1},\,\cite{ISAC2}     & --          & --          & --          & Indoor \& Outdoor  & FF               \\ 
        \cite{Indoor_ISAC}     & --          & --          & --          & Indoor  & FF              \\ 
        \cite{IIoT_noma}                                                    & \checkmark  & --          & --          & Indoor, Multi Users  & FF               \\ 
\midrule
        
        \cite{near_ris_isac}                                          & --          & \checkmark  & --          & Multi Users          & NF              \\
        \cite{near_ris_isac_2}                                          & --          & \checkmark  & --          & Multi Users, Downlink          & NF              \\

         \cite{Wang2024RIS_nf_ff_ISAC}  & -- & \checkmark & -- & Single User & NF and FF Coexistence \\

         \cite{ris_near_far_coexistence}                                     & --          & \checkmark  & --          & P2P MIMO          & NF and FF Coexistence    \\

         \cite{Luo2025RIS_ISAC_NEAR_FAR}                                              & --  & \checkmark          & --          & Multi Users, Downlink          & NF and FF Coexistence \\

         \midrule

        \cite{noma_ris_isac_1}                                              & \checkmark          & \checkmark  & --          & Multi Users, Downlink          & FF               \\

        \cite{noma_ris_isac_2}               & \checkmark  & \checkmark  & --          & Multi Users, Downlink          & FF               \\

          \cite{noma_ris_isac_3}               & \checkmark  & \checkmark  & --          & Multi Users, Downlink          & FF               \\

        \midrule
        \cite{partialNOMA2023Akhtar}                                           & \checkmark          & --          & Fixed & Single User          & FF               \\
            \cite{Semi_ISAC}                                           & \checkmark          & --          & Fixed & Multi Users, Uplink          & FF               \\
            \midrule


        This paper                                                    & \checkmark          & \checkmark  & Adaptive          &  Indoor \& Outdoor, Multi Users, Uplink          & NF and FF Coexistence    \\
        \bottomrule
    \end{tabular}
\end{table*}

\subsection{Main Contributions}
\begin{figure}[t] 
\begin{center}
\includegraphics[width=1\columnwidth]{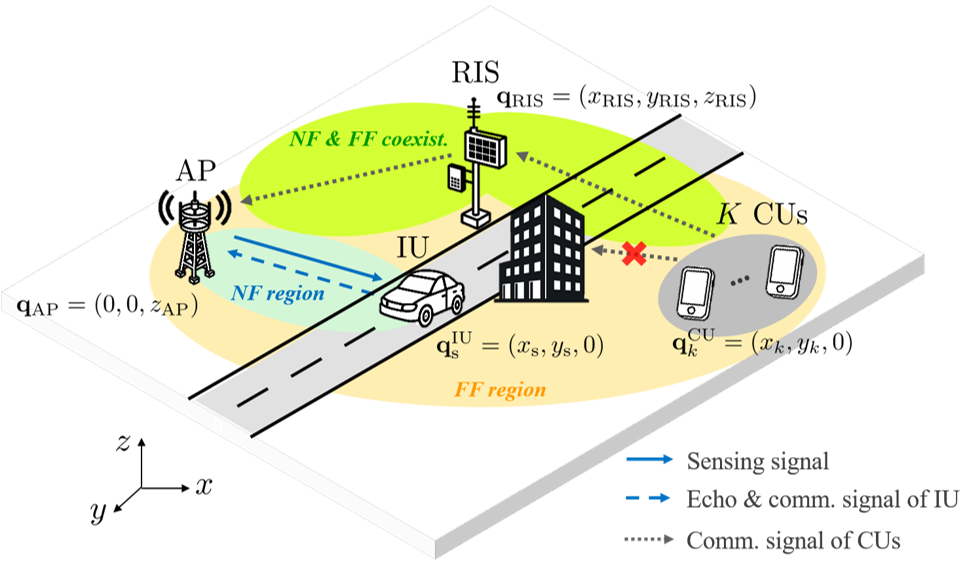}
\caption{System model of RIS-assisted ISAC systems in NF and FF coexistence. The full-duplex AP transmits a sensing signal to an IU via downlink, while at the same time it serves uplink data communication services to both IU and CUs. The IU is located within the NF region of the AP, while the CUs in the FF region operate in a hybrid NF–FF environment enabled by the RIS.}
\label{System model}
\end{center}
\end{figure}
In this paper, we propose an RIS-assisted ISAC system operating under hybrid NF and FF regions, as illustrated in Fig. \ref{System model}. Motivated by the emerging coexistence of NF and FF communications, this work introduces an adaptive bandwidth-splitting strategy for RIS-assisted ISAC systems under diverse propagation environments and heterogeneous user configurations. The key contributions are summarized as follows:
\begin{itemize}

    \item \textbf{Comprehensive review of RIS-assisted ISAC:} We provide a systematic overview of state-of-the-art RIS-assisted ISAC studies and categorize them into two primary technical directions: (i) RIS-assisted ISAC under single-field or hybrid-field scenarios, and (ii) enhanced RIS-assisted ISAC frameworks incorporating NOMA and bandwidth-splitting designs. Through this taxonomy, we highlight key frameworks, challenges and design considerations that support practical and scalable ISAC implementation in future 6G wireless networks.
    \item \textbf{Generalized ISAC framework for hybrid NF and FF conditions:} We consider an RIS-assisted ISAC architecture comprising a full-duplex access point (AP), an RIS equipped with passive reflecting elements and multiple user devices. To capture realistic propagation characteristics, we model both spherical- and planar-wave channels according to the Fresnel region \cite{Selvan2017Fraunhofer}, thereby accommodating the coexistence of NF and FF region commonly observed in practical deployments. For spectral efficiency enhancement, we propose an adaptive bandwidth-splitting mechanism that allocates an ISAC band in conjunction with a traditional sensing-only (SO) band, while employing the NOMA technique to enable prioritized decoding of multiuser communication signals followed by the processing of sensing signals.
    
    \item \textbf{Joint Beamforming and Bandwidth Splitting Optimization for RIS-enabled ISAC:} To maximize sensing accuracy, measured by the Cramér–Rao bound (CRB), we formulate a joint optimization problem that simultaneously determines the RIS phase shifts, bandwidth-splitting ratio and receive beamforming vector, subject to individual user data rate and resource constraints. An efficient algorithmic solution is developed using an alternating optimization (AO) framework combined with the semi-definite relaxation (SDR) technique. Theoretical convergence and computational complexity analyses are also provided.
    \item \textbf{Validation of adaptive bandwidth-splitting effectiveness via standard-compliant simulations:} To evaluate the performance of the proposed system, we adopt parameter configurations aligned with 3GPP standards \cite{3gpp, Zhang2025TCOM_3GPP}. Simulation results demonstrate that the proposed method significantly outperforms conventional baselines relying solely on either the ISAC or SO band, or using fixed and equal bandwidth allocation. The proposed scheme achieves superior sensing–communication trade-offs across diverse NF and FF configurations.
\end{itemize}

\subsection{Organization}
The remainder of the paper is organized as follows. Section \ref{System_Model} provides a comprehensive survey on the recent RIS-ISAC systems and introduces the set-up for the proposed RIS-assisted ISAC framework in hybrid-field scenarios and Section \ref{Communications and sensing protocols} introduces the communications and sensing protocols. Section \ref{Problem Formulation and Proposed Algorithm} formulates the CRB minimization problem and presents the proposed algorithmic solution with its convergence and complexity analyses. Section \ref{Numerical Results} provides numerical evaluations to verify the effectiveness of the proposed method. Finally, Section \ref{Conclusion} summarizes the paper and points to directions for future work.

\textit{Notation:} $\mathbb{E}[\cdot]$ denotes the expectation of the argument matrix; we reserve the superscript $A^T$ and $A^H$ for the transpose and the conjugate transpose of $A$, respectively; $\mathbf{1}_{M\times N}$ represents the all-one matrix of size $M \times N$, and $\mathrm{tr}(\cdot)$, $\mathrm{diag}(\cdot)$ and $\mathrm{rank}(\cdot)$ represent the trace, diagonal and rank function, respectively.

\section{System Model}
\label{System_Model}

In this section, we first review the state-of-the-art studies on RIS-assisted ISAC systems and then present the proposed system set-up and channel model.

\subsection{Related Works}
The recent research on RIS-assisted ISAC can be broadly categorized into two main technical directions: (i) RIS-assisted ISAC under single-field or hybrid-field scenarios, (ii) enhanced RIS-assisted ISAC incorporating NOMA and bandwidth splitting. For clarity, Table \ref{related works} summarizes the representative studies according to their primary research focus and keywords.

\subsubsection{RIS-assisted ISAC under single-field or hybrid-field scenarios}

Early studies \cite{near_ris_isac, Yoo2024ISAC, NF_RIS, near_ris_isac_2, Ruizhang_ISAC, Cho2024RIS} on single-field RIS-assisted ISAC focus on two typical propagation regions: the FF and NF regions. The FF region is characterized by planar wave propagation that forms wide beams covering a large spatial area, while the NF region exhibits spherical wavefronts that vary with distance and antenna position, enabling beam focusing and finer sensing resolution. For the NF region, the authors in \cite{near_ris_isac} and \cite{near_ris_isac_2} model the spherical wavefront within the Fresnel region of large antenna arrays and evaluated their performance in single-cell settings involving line-of-sight users and static targets. These works demonstrate that NF RIS-assisted ISAC provides enhanced resolution and beam-focusing capabilities. However, practical deployments with users at diverse link distances necessitate hybrid frameworks that can accommodate the coexistence of NF and FF propagation.

In \cite{ris_near_far_coexistence},  a piecewise NF and FF channel model is introduced for RIS-aided MIMO links under channel estimation errors. Similarly, the authors in \cite{Luo2025RIS_ISAC_NEAR_FAR} investigate a hybrid NF and FF RIS-assisted ISAC framework, in which RIS phase shifts and beamforming vectors are jointly optimized to adapt to varying propagation conditions across the Fresnel and Fraunhofer regions. Both approaches are verified to achieve the notable performance gains over conventional systems relying solely on either NF or FF assumptions, validating that hybrid designs can achieve seamless connectivity and sensing for heterogeneous users with manageable computational complexity.

\subsubsection{Enhanced RIS-assisted ISAC with NOMA and bandwidth splitting}
The NOMA has been explored increasingly to expand the design space of RIS-assisted ISAC systems. Specifically, \cite{noma_ris_isac_1} introduces a comprehensive RIS-NOMA-ISAC framework that jointly optimizes base station beamforming, NOMA power allocation and RIS phase shifts, achieving significant beam pattern and spectral efficiency improvements over conventional RIS-enabled ISAC or SO systems. In \cite{noma_ris_isac_3}, a backscatter-enabled RIS-assisted NOMA-ISAC framework with dynamic user clustering is proposed to balance communication throughput and sensing power.

The aforementioned studies \cite{noma_ris_isac_1, noma_ris_isac_2, noma_ris_isac_3}, however, allocate the entire spectrum exclusively for ISAC operation, applying NOMA within this band to improve spectral utilization. This approach may be impractical in real-world scenarios since legacy high-frequency bands are often pre-allocated for dedicated sensing purposes  \cite{Griffiths2015RadarSpectrum}. To address this issue, the authors in \cite{Bliss} introduce a joint-utilization paradigm that aggregates the existing SO band with a dedicated ISAC band, thereby improving sensing accuracy and communication throughput under spectral coexistence conditions. This concept has been further extended in \cite{partialNOMA2023Akhtar} and \cite{Semi_ISAC}, where a bandwidth-splitting ratio is introduced to partition the spectrum into communication-only, SO, and ISAC sub-bands. By employing uplink NOMA, these works effectively mitigate interference between sensing and communication signals, demonstrating improved accuracy and throughput compared to ISAC-only or OMA schemes. Nevertheless, bandwidth splitting has not yet been explored in the context of RIS-assisted ISAC systems. Existing studies \cite{Semi_ISAC, Bliss, noma_ris_isac_1} remain limited to performance analysis under predefined bandwidth-splitting configurations, without optimizing the ratio to achieve optimal sensing–communication trade-offs.

\subsubsection{Challenges and Motivation}
Despite the significant progress achieved in these research directions, several challenges remain in developing a unified and practical RIS-assisted ISAC framework. Most existing studies \cite{near_ris_isac, near_ris_isac_2, noma_ris_isac_1, noma_ris_isac_2, noma_ris_isac_3} do not explicitly account for user-specific configurations or environmental diversity, often focusing on isolated cases such as indoor or outdoor deployments. Although NF and FF coexistence has been investigated, the previous approaches rarely capture the dynamic channel transitions that occur when RIS placement or user mobility causes shifts between propagation regions. Furthermore, achieving high sensing accuracy while satisfying communication quality-of-service (QoS) requirements under NOMA-induced multiuser interference remains a difficult optimization task. To jointly manage sensing and communication resources under mixed NF and FF conditions, designing an adaptive bandwidth-splitting mechanism becomes crucial for realizing a practical and spectral-efficient RIS-assisted ISAC system.

To the best of our knowledge, as summarized in Table \ref{related works}, this is the first work that investigates an RIS-assisted ISAC system with an adaptive bandwidth-splitting strategy explicitly designed for hybrid NF and FF coexistence. By considering diverse user and RIS deployments, this study provides new insights into practical ISAC implementations and lays the groundwork for future innovations in next-generation wireless systems.

\subsection{Set-up}
We consider an RIS-assisted ISAC system under hybrid NF and FF coexistence, involving both sensing and communication functionalities, as illustrated in Fig. \ref{System model}. The AP first transmits a radar signal to a nearby target within its NF region via the downlink. Subsequently, the AP receives the reflected echo signal. At the same time, the AP serves uplink data communication services to the users, while continuously monitoring the surrounding environment to ensure reliable system operations. In order to support the uplink data communication of users in the FF region, whose direct links to the AP are obstructed by various obstacles, we deploy an RIS to establish an intermediate link, thereby enhancing link reliability and overall system performance. Specifically, depending on the placement of the RIS, the corresponding cascaded links may experience either NF or FF propagation. The AP operates in full-duplex mode, transmitting a sensing signal while concurrently receiving both corresponding echo signal and communication signals from users, where the self-interference is assumed to be perfectly suppressed \cite{SI}. The considered system consists of an AP with a uniform linear array (ULA) with $N$ elements and an antenna spacing of $d$ m, an RIS with $M$ passive transmission-reflection units with elements spacing $w$ m and single-antenna users. 

In proposed RIS-assisted ISAC scenarios as shown in Fig. \ref{System model}, the users can be categorized into two types based on their operational characteristics and distance from the AP: (i) a ISAC user (IU) as a subject to be sensing target that uses both SO and ISAC band and (ii) communication-only users (CUs) as a subject to be supported by RIS due to obstacles that use only ISAC band. Here, one IU and $K$ CUs are distinguished according to the Rayleigh distance criterion of $2D^2/\lambda$ \cite{fresnel}, where $D=(N-1)d$ is the aperture of the ULA antenna and $\lambda$ is the signal wavelength. The IU (denoted as `$\mathrm{s}$') serves both as a sensing target and a communication user, while the remaining $K$ CUs are dedicated exclusively to communication. This single-target scenario can be readily extended to multiple sensing targets through standard multiple target estimation or sequential sensing techniques \cite{Yang2024Multi_target}. For example, in outdoor environments such as urban transportation networks with dense traffic and roadside units, IU may correspond to mobile users (e.g., vehicles or UAVs) that perform sensing and communication with the AP, while CUs blocked by buildings are supported vian RISs deployed on surrounding infrastructures. In indoor scenarios, such as smart factories or logistics centers, CUs, such as stationary devices obstructed by inventories are assisted through ceiling- or wall-mounted RISs, while IU may correspond to AGVs performing joint localization and communication.

All nodes are assumed to be static, where coordinates of AP and RIS are defined as $\mathbf{q}_{\textrm{AP}}=(0\textrm{ m},0\textrm{ m},z_{\textrm{AP}}\textrm{ m})$ and $\mathbf{q}_{\text{RIS}}=(x_{\text{RIS}},y_{\text{RIS}},z_{\text{RIS}})$, respectively. The coordinates for $n$-th antenna of the AP is given by $\mathbf{q}_{\textrm{AP}(n)}=(0,0,z_\text{AP}+nd)$, for $n \in \{-N/2,-N/2+1,\ldots,N/2-1,N/2\}$, while the coordinates of $(m_1,m_2)$-th RIS element along the $x$- and $y$-axes are specified as $\mathbf{q}_{\textrm{RIS}(m_1,m_2)}=(x_\text{RIS}+mw,y_\text{RIS}+mw,z_\text{RIS})$, for $m_1, m_2 \in \{-\sqrt{M}/2,-\sqrt{M}/2+1,\ldots,\sqrt{M}/2-1,\sqrt{M}/2\}$. Note that the $n$-th antenna element of the AP and the  $(m_1,m_2)$-th element of the RIS  are denoted as $\textrm{AP}(n)$ and $\textrm{RIS}(m_1,m_2)$, respectively, and this notation is adopted throughout the paper. The coordinates of the IU and the $k$-th CU are denoted as  $\mathbf{q}_{\mathrm{s}}=(x_{\textrm{s}},y_{\mathrm{s}},0)$ and $\mathbf{q}_k=(x_k,y_k,0)$, respectively. In particular, the passive square RIS adjusts the phase of each reflecting element with unit amplitudes. The RIS serves as the cascade communication channel link between the AP and the CUs \cite{NF_RIS}. The reflection matrix is defined as $\mathbf{\Phi}=\mathrm{diag}(v_1,...,v_M)$, where $v_m=e^{j\theta_m}$, for all $m \in \mathcal{M} \triangleq \{1,2,... M\}$, and $\theta_m$ denotes the phase-shift of the $m$-th element of the RIS \cite{Ruizhang_ISAC}. Also, the measure $d_{i,j}=\lVert \mathbf{q}_{i}-\mathbf{q}_{j} \rVert$ represent the Euclidean distance between element $i$ and $j$, which is derived as $\sqrt{(x_i-x_j)^2+(y_i-y_j)^2+(z_i-z_j)^2}$. Note that $d_{\textrm{AP,s}}$ and $d_{\textrm{AP,RIS}}$ correspond to $d_{\textrm{AP}(0),\textrm{s}}$ and $d_{\textrm{AP}(0),\textrm{RIS}(0,0)}$, respectively. Similarly, $\theta_{i,j}$ denotes the center-to-center angle between element $i$ and $j$.

\subsection{ISAC Channel Model}

Throughout this paper, the channel links are modeled as either NF with spherical-wave propagation or FF with planar-wave propagation, depending on the relative location of the RIS. In the NF case, the spherical wave propagation facilitates high-resolution sensing and beam-focusing for communication \cite{NF_ISAC, NF_FF_coexistence}, while in the FF case, it is natural to serve multiple users within a common cone-shaped region using a single-beam through NOMA \cite{NF_FF_coexistence}. Note that the channel state information between nodes is assumed to be known from the previous coherence
time block via proper channel estimation methods, e.g., pilot-based estimation or radar sensing \cite{NF_ISAC}. In the following, the NF channel model of IU is firstly introduced, and then the cascaded channels of AP-RIS and RIS-CUs mixed with NF and FF are explored. 


For the sensing channel between IU and AP, we adopt the NF model \cite{Ruizhang_ISAC}, where the point target is modeled as a single scatterer with a small spatial extent, and the reflected echo signals are assumed to propagate through a single dominant path. Accordingly, the NF sensing echo channel of the single IU is modeled as \cite{NF_ISAC}
\begin{equation}
\label{sensing_channel}
\mathbf{G}_{\textrm{AP,s}}=\beta_{\textrm{s}}\mathbf{a}_{\textrm{AP,s}}\mathbf{a}_{\textrm{AP,s}}^T,
\end{equation}
where $\beta_\textrm{s}$ denotes the complex-valued channel coefficient that depends on the IU radar cross section (RCS), and $\mathbf{a}_\textrm{{AP,}s}$ denotes the NF array response vector, i.e., $\mathbf{a}_{\textrm{{AP,}s}}=\big[e^{-j\frac{2\pi}{\lambda}d_{\textrm{AP}(-N/2),\textrm{s}}},...,e^{-j\frac{2\pi}{\lambda}d_{\textrm{AP}(N/2),\textrm{s}}}\big]^T \in \mathbb{C}^{N \times 1}$. The NF communication channel vector $\mathbf{h}_{\textrm{AP,s}} \in \mathbb{C}^{N\times 1}$ between the IU and the $n$-th antenna element of the AP is expressed as
\begin{equation}
    \mathbf{h}_{\textrm{AP,s}}=\alpha_\text{AP,s}\mathbf{a}_{\textrm{AP,s}},
\end{equation}
where $\alpha_{\textrm{AP,s}}=\sqrt{{\lambda}/{(4\pi d^2_\textrm{AP,s})}}$ \cite{NF_FF_coexistence}. 
\begin{figure*}[t]
\centering
\begin{subfigure}[t]{0.35\textwidth}
    \includegraphics[width=0.95\textwidth]{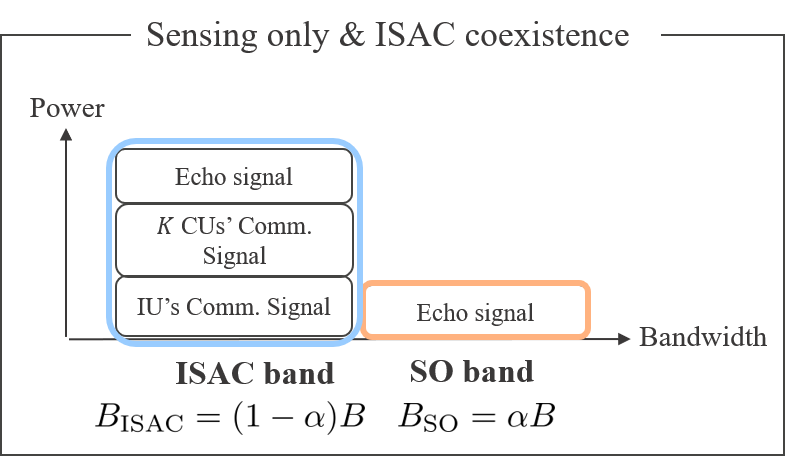}
\caption{}
\end{subfigure}
\begin{subfigure}[t]{0.8\textwidth}
    \includegraphics[width=1.1\textwidth]{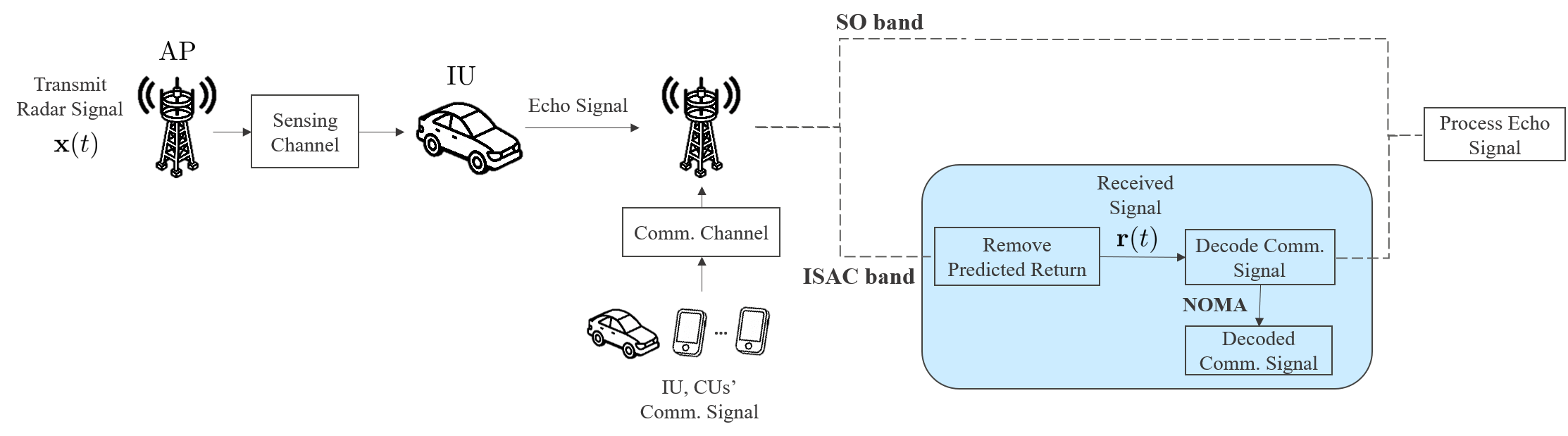}
\caption{}
\end{subfigure}
\caption{Coexistence of SO and ISAC bands. (a) Bandwidth allocation strategy. (b) Uplink and downlink procedure.}
\label{Protocol}
\end{figure*}
In the case of the cascaded channel model between the AP and the CUs via the RIS, the propagation condition is classified as either NF or FF based on the RIS location with respect to the Fresnel region. Accordingly, the channels between the RIS and the $k$-th CU, and between the AP and the RIS, are denoted by the general notations $\mathbf{h}_{\text{RIS},k}$ and $\mathbf{H}_\text{AP,RIS}$, respectively. Depending on the RIS location, $\mathbf{h}_{\text{RIS},k}$and $\mathbf{H}_\text{AP,RIS}$ are selected such that  $\mathbf{h}_{\text{RIS},k}\in \{\mathbf{h}^\text{NF}_{\text{RIS},k},\mathbf{h}^\text{FF}_{\text{RIS},k}\}$ and  $\mathbf{H}_{\text{AP,RIS}}\in \{\mathbf{H}^\text{NF}_{\text{AP,RIS}},\mathbf{H}^\text{FF}_{\text{AP,RIS}}\}$, whose details are given in the following. The NF channel vector $\mathbf{h}^\text{NF}_{\textrm{RIS}, k} \in \mathbb{C}^{M \times 1}$ between the CU $k$ and the RIS is defined as \cite{NF_ISAC}
\begin{equation}
\begin{aligned}
    \mathbf{h}^\text{NF}_{\textrm{RIS}, k}=\alpha_{\text{RIS},k}\big[&e^{-j\frac{2\pi}{\lambda}d_{\textrm{RIS}(-\sqrt{M}/2,-\sqrt{M}/2),k}},...,\\&e^{-j\frac{2\pi}{\lambda}d_{\textrm{RIS}(\sqrt{M}/2,\sqrt{M}/2),k}}\big]^T, \quad \forall k \in \mathcal{K},
\end{aligned}
\end{equation}
where $\alpha_{\text{RIS},k}=\sqrt{{\lambda}/{(4\pi d^2_{\textrm{RIS},k})}}$. In addition, the NF channel matrix $\mathbf{H}^\text{NF}_\textrm{AP,RIS}\in \mathbb{C}^{M\times N}$ from the RIS and the AP is given as
\begin{equation}
\begin{aligned}
\mathbf{H}^\text{NF}_\textrm{AP,RIS}=\big[ \mathbf{h}^\text{NF}_{\text{AP,RIS}(-\sqrt{M}/2,-\sqrt{M}/2)}&,\ldots,\mathbf{h}^\text{NF}_{\text{AP,RIS}({m_1,m_2})},\ldots,\\&\mathbf{h}^\text{NF}_{\text{AP,RIS}(\sqrt{M}/2,\sqrt{M}/2)}\big]^T,
\end{aligned}
   \end{equation}
   where $\mathbf{h}_{\text{AP,RIS}({m_1,m_2})}=\alpha_\text{AP,RIS}\big[e^{-j\frac{2\pi}{\lambda}d_{\textrm{AP}(-N/2),\textrm{RIS}(m_1,m_2)}},\ldots,\\ e^{-j\frac{2\pi}{\lambda}d_{\textrm{AP}(N/2),\textrm{RIS}(m_1,m_2)}}\big]^T$ and $\alpha_\text{AP,RIS} \hspace{-3pt} =\hspace{-3pt}\sqrt{{\lambda}/{(4\pi d_{\textrm{AP,RIS}}^2)}}$. On the other hand, when the distances between nodes exceed the Fresnel distance, a FF channel model based on the planar-wave assumption is considered. In this case, the FF channel vector from $k$-th CU to the RIS is given by \cite{NF_FF_coexistence}
\begin{equation}
\begin{aligned}
\mathbf{h}_{\text{RIS},k}^{\text{FF}}=\alpha_{\text{RIS},k} \exp \big( -j\frac{2\pi}{\lambda}d_{\text{RIS},k}\big)\mathbf{b}_{\text{RIS},k},
\end{aligned}
\end{equation}
where $\mathbf{b}_{\text{RIS},k}=\big[1,e^{-j\frac{2\pi w}{\lambda}\sin \theta_{\text{RIS},k}},\ldots,e^{-j\frac{2\pi w}{\lambda}(M-1)\sin \theta_{\text{RIS},k}}\big]^T$ is the steering vector between RIS and CU $k$.  Similarly, the FF channel matrix from the RIS and the AP is represented by \begin{equation}
\mathbf{H}_{\text{AP,RIS}}^{\text{FF}}=\alpha_\text{AP,RIS} \exp \big(-j\frac{\lambda}{2\pi}d_\text{AP,RIS} \big)\mathbf{b}_\text{AP,RIS}\cdot\mathbf{b}^H_\text{RIS,AP},
\end{equation}
where $\mathbf{b}_{\text{AP,RIS}}=\big[1,e^{-j\frac{2\pi d}{\lambda}\sin \theta_{\text{AP,RIS}}},\ldots,e^{-j\frac{2\pi d}{\lambda}(N-1)\sin \theta_{\text{AP,RIS}}}\big]^T$ and $\mathbf{b}_{\text{RIS,AP}}=\big[1,e^{-j\frac{2\pi w}{\lambda}\sin \theta_{\text{AP,RIS}}},\ldots,e^{-j\frac{2\pi w}{\lambda}(M-1)\sin \theta_{\text{AP,RIS}}}\big]^T$. 


\section{Communications and Sensing Protocols}
\label{Communications and sensing protocols}
In this section, we develop the protocol for the coexistence of SO and ISAC bands, and the corresponding NOMA-based communication and radar sensing model.
\subsection{Adaptive Bandwidth Splitting Protocol}

We adopt a protocol incorporating both SO band and ISAC band, as depicted in Fig. \ref{Protocol}(a). From a practical perspective, since the majority of high-frequency spectral resources are already assigned to radar sensing services \cite{Semi_ISAC}, allocating exclusive bandwidth for ISAC is challenging. To address this limitation, we adopt a protocol that allocates a portion of the existing SO band within the mmWave and sub-THz spectrum to ISAC operations, instead of relying solely on dedicated ISAC bands \cite{Bliss}. Specifically, the total bandwidth $B$ with the center frequency $f$ is partitioned into the SO band and the ISAC band\footnote{This work assumes ideal band separation without considering guard bands or inter-band interference for analytical tractability. Nevertheless, incorporating practical guard bands would not significantly affect the applicability of the proposed bandwidth splitting framework, which is left as an interesting direction for future work.} according to a bandwidth splitting ratio $\alpha$ , yielding
\begin{equation}
    B_\textrm{SO}=\alpha B \textrm{ and } B_\textrm{ISAC}=(1-\alpha)B.
\end{equation}

Next, we introduce the detailed adaptive bandwidth-splitting protocol that operates throughout the overall process, in which the AP transmits radar signals and receives both echo signals and uplink communication signals from the users, as illustrated in Fig. \ref{Protocol}(b). Firstly, the AP transmits downlink radar signals to the IU. During the uplink procedure, the AP receives echo signals from IU across both SO and ISAC bands, while the uplink communication signals from IU and CUs are received solely in the ISAC band. Since the ISAC and SO bands are exploited concurrently, the overall operation mandates a dedicated bandwidth-splitting protocol to effectively coordinate the use of two spectra.
\begin{figure*}[!t]
\begin{equation}
\label{sinr_s}
    \gamma_\mathrm{s}(\alpha, \mathbf{f},\boldsymbol{\Phi}) = \dfrac{|\mathbf{f}^H\mathbf{h}_{\textrm{AP,s}}|^2p_\text{s}}{|\mathbf{f}^H\sum_{k=1}^K \bar{\mathbf{h}}_{\textrm{AP},k}(\mathbf{\Phi})|^2p_k+|\mathbf{f}^H\mathbf{a}_{\textrm{AP,s}}|^2\beta_\text{s}^2\eta_\text{ISAC}p_\textrm{ISAC}+(1-\alpha)\sigma^2} \textrm{ and}\tag{12}
\end{equation}
\begin{equation}
\label{sinr_k}
    \gamma_{k}(\alpha, \mathbf{f},\boldsymbol{\Phi}) = \dfrac{|{\mathbf{f}}^H\bar{\mathbf{h}}_{\textrm{AP},k}(\mathbf{\Phi})|^2p_k}{\sum_{i \in \mathcal{K},i \neq k}|\mathbf{f}^H \bar{\mathbf{h}}_{\textrm{AP},i}(\mathbf{\Phi})|^2p_i+|\mathbf{f}^H\mathbf{a}_{\textrm{{AP,}s}}|^2\beta_\text{s}^2\eta_\text{ISAC}p_\textrm{ISAC}+(1-\alpha)\sigma^2}, \quad \forall k \in \mathcal{K}.\tag{13}
\end{equation}
\hrulefill
\vspace{-0.5cm}
\end{figure*}
The radar signals $x_\text{SO}(t)$ and $x_\text{ISAC}(t)$ of SO and ISAC band are transmitted from AP toward the IU over both the SO and ISAC bands, with the linear transmit beamforming $\mathbf{w}_\text{SO} \in \mathbb{C}^{N \times 1}$ and $\mathbf{w}_\text{ISAC}  \in \mathbb{C}^{N \times 1}$. Accordingly, the transmit signal of AP can be expressed as
\begin{equation}
    \mathbf{x}(t)=\mathbf{w}_\textrm{SO}{x}_\textrm{SO}(t)+\mathbf{w}_\textrm{ISAC}{x}_\textrm{ISAC}(t),
    \label{SO_ISAC_tx}
\end{equation}
where radar signals $x_\textrm{SO}(t)$ and $ x_\textrm{ISAC}(t)$ have unit average power, i.e., $\mathbb{E}[|x_\textrm{SO}(t)|^2]=\mathbb{E}[|x_\textrm{ISAC}(t)|^2]=1$. In order to maximize the sensing accuracy, the AP directly needs to adjust the beam towards the IU \cite{qingqingwu_ISAC}, which allows it to have
\begin{equation}
\mathbf{w}_\textrm{SO}=\sqrt{\dfrac{p_\textrm{SO}}{N}}\frac{({\mathbf{a}_{\textrm{{AP,}s}}^T})^H}{\lVert{\mathbf{a}_{\textrm{{AP,}s}}^T}\rVert} \textrm{ and } \mathbf{w}_\textrm{ISAC}=\sqrt{\dfrac{p_\textrm{ISAC}}{N}}\frac{({\mathbf{a}_{\textrm{{AP,}s}}^T})^H}{\lVert{\mathbf{a}_{\textrm{{AP,}s}}^T}\rVert},
\end{equation}
where 
\begin{equation}
    p_\textrm{SO}=\alpha B \rho_\textrm{SO} \textrm{ and } p_\textrm{ISAC}=(1-\alpha) B \rho_\textrm{ISAC},
\end{equation}
and $\rho_\textrm{SO}$ and $\rho_\textrm{ISAC}$ are the power spectral densities of the SO band and ISAC band, respectively. The radar signal propagates through the sensing channel $\mathbf{G}_\text{AP,s}$ in (\ref{sensing_channel}), and its corresponding echo signal is received at the AP along with the uplink communication signals from the users.

\subsection{NOMA-based Communication Model}
To efficiently support joint sensing and communication services for multiple users, NOMA is adopted in the uplink procedure, where SIC is leveraged to extract the echo and communication signals of users from the superimposed received signals \cite{Semi_ISAC}. In general, the communication signals convey more information than radar pulses and are therefore assigned higher decoding priority than echo signals. However, assuming that the AP can estimate the predicted time delay $\tau_{pre}$ of the echo signal based on prior observations \cite{Bliss,Semi_ISAC}, the interference from communication signals can be effectively mitigated by subtracting the predicted return from the received superimposed signals. 

In the ISAC band, the received signal $\mathbf{r}(t)$ at AP is written as
\begin{equation}
\begin{aligned}
&\mathbf{r}(t)=\hspace{-8pt}\underbrace{\mathbf{h}_{\textrm{AP,s}}\sqrt{p_{\textrm{s}}}s_{\textrm{s}}(t)}_{\textrm{{IU}'s communication signal}}\hspace{-7pt}+\hspace{-3pt}\underbrace{\sum\limits_{k=1}^K \bar{\mathbf{h}}_{\textrm{AP},k}(\mathbf{\Phi})\sqrt{p_k}s_{k}(t)}_{\textrm{$K$ CUs' communication signal}}\\&+\underbrace{\mathbf{G}_{\textrm{AP,s}}\big[\mathbf{w}_\text{ISAC}x_\text{ISAC}(t-\tau)-\mathbf{w}_\text{ISAC}x_\text{ISAC}(t-\tau_{pre})\big]}_{\textrm{Echo signal with reduced power level}}+\mathbf{z}(t),
\end{aligned}
\end{equation}
where $\tau$ is the time delay fluctuation to the IU and it follows a Gaussian distribution with the variance of $\sigma_{\tau}^2=\mathbb{E}[|\tau-\tau_{pre}|^2]$. The transmit power of the {IU} and CU $k$ are denoted by $p_\text{s}$ and $p_k$, respectively, $\mathbf{z}(t) \sim \mathcal{CN}(0,(1-\alpha)\sigma^2\mathbf{I}_N)$ represents the additive white Gaussian noise (AWGN) at the AP, and $\bar{\mathbf{h}}_{\textrm{AP},k}(\mathbf{\Phi})\triangleq \mathbf{H}_\textrm{AP,RIS}^{H}\boldsymbol{\Phi}\mathbf{h}_{\textrm{RIS}, k}$ represents combined channel vector from the CU $k$ to the AP vian RIS. To receive the high-quality signals at the full-duplex AP, we employ a receive beamforming $\mathbf{f}$, resulting in the received signal $y(t)=\mathbf{f}^H\mathbf{r}(t)$. According to the uplink NOMA procedure \cite{Semi_ISAC}, the echo signal reflected from the IU is detected first. Accordingly, the signal-to-interference-plus-noise ratio (SINR) of the echo signal is given by (\ref{sinr_s}), where $\eta_\text{ISAC}=(2\pi)^2B_\text{ISAC}^2 \sigma_{\tau}^2/12$ denotes the average reduced power levels of the echo in the ISAC band \cite{Semi_ISAC}. Although the communication signal from the IU is completely canceled under the assumption of perfect SIC, the AP still sequentially decodes the uplink signals from the $K$ CUs in the presence of residual interference from other CUs and echo signals with reduced power levels. To this end, the SINR of the $k$-th CU can be derived as in (\ref{sinr_k}).

\subsection{Radar Sensing Model}
Since the sensing signal is processed across both the SO and ISAC bands, the objective of the overall system is to maximize the sensing performance of the time-delay estimator. To this end, we motivate the coexistence of the SO band and the ISAC band by minimizing the CRB, which is derived based on the residual sensing signal after decoding the communication signals.

After decoding the communication signals from the users, they can be successfully removed via SIC, thereby enabling the extraction of the radar echo signal at its original power level without communication interference. However, the residual interference from the communication signals may remain during echo signal processing. Accordingly, the radar signal transmitted by the AP in (\ref{SO_ISAC_tx}) is reflected by the IU, and corresponding echo signals with original power level are received at the AP over both SO and ISAC bands, which can be expressed as
\setcounter{equation}{14}
\begin{equation}
\label{crb_signal}
\begin{split}
    &{y}_\textrm{s}(t)=\frac{\beta_\textrm{s}\mathbf{f}^H\mathbf{a}_\textrm{AP,s}}{\sqrt{N}}\big(\big[\underbrace{\sqrt{p_\text{SO}}{x}_\textrm{SO}(t-\tau)}_\text{Echo signal in SO band}+z_\text{SO}(t)\big]+\\ &\big[\underbrace{\sqrt{p_\text{ISAC}}{x}_\textrm{ISAC}(t-\tau)}_\text{Echo signal in ISAC band}+\underbrace{b\sqrt{p_\mathrm{res} }s_\mathrm{res}(t)}_\text{Residual communication signal}+z_\text{ISAC}(t)\big]\big)\\
    &=ax(t-\tau)+{{b}\sqrt{p_\mathrm{res} }s_\mathrm{res}(t)}+{\tilde{z}}(t),
\end{split}
    \end{equation}
where we have defined $x(t-\tau)=\sqrt{p_\text{SO}}{x}_\textrm{SO}(t-\tau)+\sqrt{p_\text{ISAC}}{x}_\textrm{ISAC}(t-\tau)$, $b\sqrt{p_\mathrm{res}}s_\mathrm{res}(t)$ is the residual communication signal remaining after SIC, $z_\text{SO}\sim \mathcal{CN}(0,\alpha{\sigma}^2)$ and $z_\text{ISAC}\sim \mathcal{CN}(0,(1-\alpha){\sigma}^2)$ are AWGN in SO band and ISAC band, respectively, and $\tilde{z}(t)$ is the sum of both bands noise with the variance $\sigma^2$ of AWGN. 

Based on the received signal in (\ref{crb_signal}), we evaluate the CRB as the sensing performance metric, which provides the lower bound for the variance of time-delay estimator \cite{crb}. Let us denote $\theta=\tau$ as the parameter to estimate. The received signal $y_\text{s}(t)$ is modeled as a distribution $y_\text{s}(t) \sim \mathcal{CN}\big(ax(t-\tau)+b\sqrt{p_\mathrm{res}}s_\mathrm{res}(t),\sigma^2\big)$, with the corresponding probability density function is given by $p(y_\text{s}(t),\theta)={e^{-{\lVert y_\mathrm{s}(t)-ax(t-\tau)-b\sqrt{p_\mathrm{res}}s_\mathrm{res}(t)\rVert^2}/{\sigma^2}}}/({\pi \sigma^2})$. According to \cite{Bliss}, the Fisher information matrix (FIM) for the estimation is derived as
\begin{equation}
\label{crb_derive}
\begin{split}
    &{J}=\mathbb{E}_{t}\big[f\big(\theta;z(t))f^*(\theta;z(t)\big)\big]\\
    &\hspace{-4pt}=\dfrac{2\lVert a\rVert^2\mathbb{E}\big[\lVert\sqrt{p_\text{SO}}x'_\text{SO}(t-\tau)\hspace{-3pt}+\hspace{-3pt}\sqrt{p_\text{ISAC}}x'_\text{ISAC}(t-\tau)\rVert^2\big]}{\sigma^2},
\end{split}
\end{equation}
where $f(\theta;y_\mathrm{s}(t))={\partial}\{\log p(y_\text{s}(t);\theta)\}/{\partial\theta}$ is the score function and $x'(t-\tau)=\partial x(t-\tau)/\partial \tau$. Subsequently, by applying Parseval's theorem \cite{bliss2013parseval} to transition from the time domain to the frequency domain, and exploiting the time shift and differentiation properties of the Fourier transform, the FIM can be expressed as
\begin{equation}
\label{FIM_eq}
     \begin{aligned}
     &J(\alpha,\mathbf{f})=\dfrac{8\pi^2 \beta_{\textrm{s}}^2 | \mathbf{f}^H\mathbf{a}_{\textrm{AP,s}} |^2 T B}{3N\sigma^2}\times \\
     &\bigg[\alpha\rho_\textrm{SO}\big(\big(B_O-\frac{B}{2}+\alpha B\big)^3-\big(B_O-\frac{B}{2}\big)^3\big)\\&-(1-\alpha)\rho_\textrm{ISAC}\big(\big(B_O-\frac{B}{2}+\alpha B\big)^3-\big(B_O+\frac{B}{2}\big)^3\big)\bigg],
     \end{aligned}
\end{equation}
where $B_O$ is a free parameter and is determined based on the reduced FIM as presented in \cite{Bliss}. The details of the derivation of (\ref{FIM_eq}) are given in the Appendix A. Consequently, based on the FIM, the CRB for estimating the time-delay $\tau$ is given by \begin{equation}
    \mathrm{CRB}(\alpha,\mathbf{f})=\frac{1}{J(\alpha,\mathbf{f})}.
\end{equation}

\section{Problem Formulation and Proposed Algorithm}
\label{Problem Formulation and Proposed Algorithm}
\subsection{Problem Formulation}
We aim to minimize the CRB of the time-delay estimator by jointly optimizing the receive beamforming vector $\mathbf{f}$, the bandwidth splitting ratio $\alpha$ and the RIS phase shift $\mathbf{\Phi}$. To this end, we formulate the following optimization problem:
\begin{subequations}\label{p1}
\begin{eqnarray}
&&\hspace{-1.3cm}\textrm{(P1)}: \min_{\substack{\alpha,\mathbf{f},\boldsymbol{\Phi}}} \,\,\textrm{CRB}(\alpha,\mathbf{f})\\[2pt]
&&\hspace{-1.2cm}\text{s.t.}\hspace{+0.4cm} \lVert\mathbf{f}\rVert=1, \\[2pt]
&&\hspace{-0.4cm} (1-\alpha)B\log_2 (1+\gamma_\mathrm{s}(\alpha, \mathbf{f},\boldsymbol{\Phi})) \geq R_\mathrm{s}^{th}, \\[2pt]
&&\hspace{-0.4cm} (1-\alpha)B\log_2 (1+\gamma_{k}(\alpha, \mathbf{f},\boldsymbol{\Phi})) \geq R_{k}^{th},  \forall k \in \mathcal{K}, \\[2pt]
&&\hspace{-0.4cm} |{v}_{m}|=1, \forall m \in \mathcal{M},\\[2pt]
&&\hspace{-0.4cm} 0 \leq \alpha \leq 1,
\end{eqnarray}
\end{subequations}
where (\ref{p1}b) ensures that the receive beamforming vector satisfies the unit modulus constraint, (\ref{p1}c) and (\ref{p1}d) guarantee the data rate of the IU and $k$-th CU, respectively, (\ref{p1}e) represents the constraint for each element of RIS phase shift, and (\ref{p1}f) is the constraint on the bandwidth division ratio made under the assumption that ISAC band and SO band coexist.

The optimization problem (P1) is inherently non-convex due to the non-convexity of the objective function (\ref{p1}a) and the coupled variables in constraints (\ref{p1}c) and (\ref{p1}d), which involve the receive beamforming vector $\mathbf{f}$ and the RIS phase shift matrix $\mathbf{\Phi}$. Additionally, the unit-modulus constraint in (\ref{p1}e) further contributes to the non-convexity of the feasible set. To address these challenges, we propose an Algorithm 1 based on the AO framework combined with SDR techniques that effectively obtains the local optimal solutions of problem (P1).

\subsection{Optimization of RIS-assisted ISAC}
In this section, the details of proposed AO-based algorithm 1 are described as below. Specifically, given the RIS phase shift matrix $\mathbf{\Phi}$ and the bandwidth splitting ratio $\alpha$, the receive beamforming vector $\mathbf{f}$ is optimized by introducing slack variables and applying an SDR technique. Next, for the given $\mathbf{f}$ and $\mathbf{\Phi}$, the bandwidth splitting ratio $\alpha$ is optimized based on the general convex optimization method. Lastly, for the given $\mathbf{f}$ and $\alpha$, the RIS phase shift matrix $\mathbf{\Phi}$ is optimized by employing a similar strategy using slack variables and SDR.

\subsubsection{Optimization of Receive Beamforming $\mathbf{f}$}
For the given $\mathbf{\Phi}$ and $\alpha$, the receive beamforming of problem (P1) can be rewritten as
\begin{subequations}\label{p2-1}
\begin{eqnarray}
&&\hspace{-1.3cm}\textrm{(P2-1)}: \min_{\substack{\mathbf{f}}} \,\,\textrm{CRB}(\mathbf{f})\\[2pt]
&&\hspace{-1.2cm}\text{s.t.}\hspace{+0.4cm} \textrm{(\ref{p1}b)-(\ref{p1}d)}. 
\end{eqnarray}
\end{subequations}
In (P2-1), the constraints (\ref{p1}c) and (\ref{p1}d) are non-convex since the SINR involved with the optimization variable $\mathbf{f}$ is in the logarithm term. Moreover, the objective function (\ref{p2-1}a) is not guaranteed to be convex, as the squared norm of the optimization variable appears in the denominator. To this end, we first introduce a slack variable $\mathbf{F}\triangleq \mathbf{f}\mathbf{f}^H$, which needs to satisfy $\mathbf{F}\succeq 0$ and $\mathrm{rank}(\mathbf{F})=1$. Consequently, constraint (\ref{p1}b) can be rewritten as $\mathrm{tr}(\mathbf{F})=1$. By substituting the original optimization variable $\mathbf{f}$ with the slack variable $\mathbf{F}$ and relaxing the rank-one constraint, the problem (P2-1) can be reformulated as
\begin{subequations}\label{p2-2}
\begin{eqnarray}
&&\hspace{-1.5cm}\textrm{(P2-2)}: \min_{\substack{\mathbf{F}}} \,\,{A_{\alpha}^{-1}}{\mathrm{tr}(\mathbf{F}\mathbf{a}_\text{AP,s}\mathbf{a}_\text{AP,s}^H)}^{-1}\\[2pt]
&&\hspace{-1.4cm}\text{s.t.}\hspace{+0.4cm} \begin{aligned}
&p_\text{s}(\mathbf{h}_{\textrm{AP,s}}^H\mathbf{F}\mathbf{h}_{\textrm{AP,s}}) \geq C_{\alpha}\big(p_k(\sum_{k=1}^K\mathbf{\bar{h}}_{\textrm{AP},k}^H)\mathbf{F}(\sum_{k=1}^K\mathbf{\bar{h}}_{\textrm{AP},k})\\&+{\beta_\text{s}^2\eta_\text{ISAC}p_\textrm{ISAC}}(\mathbf{a}_\text{AP,s}^H\mathbf{F}\mathbf{a}_\text{AP,s})+(1-\alpha)\sigma^2\big),\end{aligned} \\[2pt]
&&\hspace{-0.4cm} \begin{aligned}
&p_k\mathbf{\bar{h}}_{\textrm{AP},k}^H\mathbf{F}\mathbf{\bar{h}}_{\textrm{AP},k} \geq D_{\alpha}\big(\sum_{i \in \mathcal{K},i \neq k}p_i\mathbf{\bar{h}}_{\textrm{AP},i}^H\mathbf{F}\mathbf{\bar{h}}_{\textrm{AP},i}\\&+{\beta_\text{s}^2\eta_\text{ISAC}p_\textrm{ISAC}}(\mathbf{a}_\text{AP,s}^H\mathbf{F}\mathbf{a}_\text{AP,s})\\&\quad\quad\quad\quad\quad\quad+(1-\alpha)\sigma^2\big), \forall k \in \mathcal{K},\end{aligned} \\[2pt]
&&\hspace{-0.4cm} \mathbf{F} \succeq 0, \\[2pt]
&&\hspace{-0.4cm} \mathrm{tr}(\mathbf{F})=1,
\end{eqnarray}
\end{subequations}
where $A_{\alpha}={8\pi^2 \beta_{\textrm{s}}^2 T B}\big[\alpha\rho_\textrm{SO}\big((B_O-{B}/{2}+\alpha B\big)^3-\big(B_O-{B}/{2})^3\big)-(1-\alpha)\rho_\textrm{ISAC}\big((B_O-{B}/{2}+\alpha B)^3-\big(B_O+{B}/{2})^3\big)\big]/({3N\sigma^2})$, $C_{\alpha}=2^{{R_\text{s}^{th}}/{(1-\alpha)B}}-1$ and $D_{\alpha}=2^{{R_k^{th}}/{(1-\alpha)B}}-1$. Next, we drop the rank-one constraint (\ref{p2-2}e) and have a convex semi-definite program (SDP) that can be optimally solved by convex solvers, e.g., CVX \cite{cvx}. After obtaining the optimal solution $\mathbf{F}$, the rank-one solution can be efficiently recovered by taking the principal eigenvector, which is then normalized to satisfy the unit-modulus constraint. The reconstructed solution is feasible for the considered constraints, and if not, a randomization refinement is applied to ensure feasibility with near-optimal performance \cite{Sankar2024Rank1}.

\subsubsection{Optimization of Bandwidth Splitting Ratio $\alpha$} For tractability, to facilitate the optimization of the bandwidth splitting ratio $\alpha$ with the given $\mathbf{f}$ and $\mathbf{\Phi}$, we transform the CRB minimization problem of (P1) into an FIM maximization problem:
\begin{subequations}\label{p3}
\begin{eqnarray}
&&\hspace{-1.3cm}\textrm{(P3)}: \max_{\substack{\alpha}} \,\,J(\alpha),\\[2pt]
&&\hspace{-1.2cm}\text{s.t.}\hspace{+0.4cm} \textrm{(\ref{p1}c), (\ref{p1}d), (\ref{p1}f)}.  
\end{eqnarray}
\end{subequations}
In (P3), $J(\alpha)$ can be either concave or convex to the bandwidth splitting ratio $\alpha$. To address this difficulty, we introduce the following lemma.\\
\textit{\underline{Lemma} 1:} The FIM function in (\ref{FIM_eq}) is convex within the range of $0 \leq \alpha \leq 1$.\\
\textit{Proof}: See Appendix B.\\
By using Lemma 1 that $J(\alpha)$ is convex with respect to $\alpha$ for $0 \leq \alpha \leq 1$, the maximum value of $J(\alpha)$ can be achieved at the boundaries of $\alpha=0$ or $\alpha=1$ under the constraints in (\ref{p1}c) and (\ref{p1}d). Therefore, we obtain the optimal value $\alpha^*$ via a one-dimensional search over the interval $[0, 1]$, e.g., via either a grid search or the bisection method \cite{gridsearch}. Specifically, we discretize $\alpha$ interval with a step size of $\nu$ within $[0,1]$ and select the set of $\alpha$ that satisfy constraints (\ref{p1}c) and (\ref{p1}d). Among them, the optimal $\alpha$ to obtain the minimal $\mathrm{CRB}$ of problem (P3) is defined as 
\begin{equation}
    \alpha^*=\mathrm{argmin}_{\alpha \in \mathcal{A}}\mathrm{CRB}(\alpha),
\end{equation}
where $\mathcal{A}=\{\alpha\,\vert\,(1-\alpha)B\log_2 (1+\gamma_\mathrm{s}(\mathbf{f},\boldsymbol{\Phi})) \geq R_\mathrm{s}^{th} \textrm{ and } (1-\alpha)B\log_2 (1+\gamma_{k}(\mathbf{f},\boldsymbol{\Phi})) \geq R_{k}^{th}$, for $k \in \mathcal{K}$.
\subsubsection{Optimization of RIS Phase Shift $\mathbf{\Phi}$} For the given $\mathbf{f}$ and $\mathbf{\Phi}$, the optimization problem (P1) of RIS phase shift can be reformulated as
\begin{subequations}\label{p4-1}
\begin{eqnarray}
&&\hspace{-1.3cm}\textrm{(P4-1)}: \textrm{Find}\hspace{0.2cm} \boldsymbol{\Phi}\\[2pt]
&&\hspace{-1.2cm}\text{s.t.}\hspace{+0.4cm} \textrm{(\ref{p1}c)-(\ref{p1}e)}.
\end{eqnarray}
\end{subequations}
The (P4-1) cannot be directly reformulated as a second-order cone program (SOCP) optimization due to the presence of non-convex unit-modulus constraints in (\ref{p1}e). Nevertheless, the above problem resembles that of the conventional multi-user RIS beamforming optimization problem \cite{Ruizhang_ISAC}. In particular, since the constraints (\ref{p1}c) and (\ref{p1}d) can be reformulated into quadratic constraints by introducing $\mathbf{r}_k=\mathrm{diag}(\mathbf{f}^H\mathbf{H}_\textrm{AP,RIS}^H)\mathbf{h}_{\textrm{RIS},k}$, the problem (P4-1) is then transformed into
\begin{subequations}\label{p4-2}
\begin{eqnarray}
&&\hspace{-1.4cm}\textrm{(P4-2)}: \textrm{Find}\hspace{0.2cm} \boldsymbol{\Phi}\\[2pt]
&&\hspace{-1.3cm}\text{s.t.}\hspace{+0.4cm}  \begin{aligned}
|\mathbf{f}^H\mathbf{h}_{\textrm{AP,s}}|^2&p_\text{s} \geq C_{\alpha}\big(p_k\mathbf{{v}}^H\mathbf{S}\mathbf{{v}}+\\&|\mathbf{f}^H\mathbf{a}_\text{AP,s}|^2\beta_\text{s}^2\eta_\text{ISAC}p_\textrm{ISAC}+(1-\alpha)\sigma^2\big),    
\end{aligned} \\[2pt]
&&\hspace{-0.4cm} \begin{aligned}
p_k&\mathbf{{v}}^H\mathbf{R}_{k}\mathbf{{v}} \geq D_{\alpha}\big(\sum_{i \in \mathcal{K},i \neq k}p_i\mathbf{{v}}^H\mathbf{R}_{i}\mathbf{{v}}+\\&|\mathbf{f}^H\mathbf{a}_\text{AP,s}|^2\beta_\text{s}^2\eta_\text{ISAC}p_\textrm{ISAC}+(1-\alpha)\sigma^2\big), \forall k \in \mathcal{K},
\end{aligned} \\[2pt]
&&\hspace{-0.4cm} |{v}_{m}|^2=1,  m \in \{1,...,M\},
\end{eqnarray}
\end{subequations}
where $\mathbf{S}=\sum_{k=1}^K \mathbf{r}_k (\sum_{k=1}^K \mathbf{r}_k)^H$ and $\mathbf{R}_{k}=\mathbf{r}_k\mathbf{r}_k^H$. With the fact of ${\mathbf{v}}^H\mathbf{S}{\mathbf{v}}=\mathrm{tr}(\mathbf{S}{\mathbf{v}}{\mathbf{v}}^H)$ and ${\mathbf{v}}^H\mathbf{R}_i{\mathbf{v}}=\mathrm{tr}(\mathbf{R}_i{\mathbf{v}}{\mathbf{v}}^H)$ and defining $\mathbf{V}\triangleq {\mathbf{v}}{\mathbf{v}}^H$, the problem (P4-2) becomes convex by relaxing the rank-one constraint. Moreover, to achieve the better convergence solution, we generally transform the problem (P4-2) with the explicit objective function of obtaining a more efficient solution of RIS phase shift \cite{Ruizhang_ISAC}. By introducing slack variables $\delta_s$ and $\delta_k$ for all $k \in \mathcal{K}$, which can be interpreted as the residual SINR of IU and CU $k$ in phase shift optimization, the problem (P4-2) can be transformed into
\begin{subequations}\label{p4-3}
\begin{eqnarray}
&&\hspace{-0.9cm}\textrm{(P4-3)}: \max_{\substack{\mathbf{\mathbf{V},\delta_{\textrm{s}},\{\delta_\mathit{k} \}}}}\hspace{0.2cm} \delta_{\textrm{s}}+\sum_{k\in\mathcal{K}}\delta_k\\[2pt]
&&\hspace{-1cm}\text{s.t.}\hspace{+0.2cm}  \begin{aligned}
|\mathbf{f}^H\mathbf{h}_{\textrm{AP,s}}&|^2p_\text{s} \geq C_{\alpha}\big(p_k\mathrm{tr}(\mathbf{S}\mathbf{V})+\\&|\mathbf{f}^H\mathbf{a}_\text{AP,s}|^2\beta_\text{s}^2\eta_\text{ISAC}p_\textrm{ISAC}+(1-\alpha)\sigma^2\big)+ \delta_{\textrm{s}}, 
\end{aligned} \\[2pt]
&&\hspace{-0.4cm} 
\begin{aligned}
&p_k\mathrm{tr}(\mathbf{R}_{k}\mathbf{V})\geq D_{\alpha}\big(\sum_{i \in \mathcal{K},i \neq k}p_i\mathrm{tr}(\mathbf{R}_{i}\mathbf{V})\\&+|\mathbf{f}^H\mathbf{a}_\text{AP,s}|^2\beta_\text{s}^2\eta_\text{ISAC}p_\textrm{ISAC}+(1-\alpha)\sigma^2\big)+ \delta_k,\forall k \in \mathcal{K},
\end{aligned} \\[2pt]
&&\hspace{-0.4cm} \mathbf{V}_{m,m}=1,  m \in \{1,...,M\},\\[2pt]
&&\hspace{-0.4cm} \mathbf{V} \succeq 0.
\end{eqnarray}
\end{subequations}
Note that (P4-3) can be solved by using the SDR technique \cite{Ruizhang_ISAC} and can be optimally solved efficiently by optimization tools such as CVX \cite{cvx}. The rank-one solution can be similarly reconstructed from the relaxed matrix to satisfy the unit-modulus constraint, following the same SDR-based recovery procedure described earlier.
\begin{algorithm}[t]
	\caption{Proposed Algorithm for (P1).}
	\textbf{Initialize:} $\{\mathbf{\Phi}, \alpha, \mathbf{f}\}$ that satisfy the constraints (\ref{p1}b)-(\ref{p1}f), Calculate the objective function in (\ref{p1}a) with the initialized variables and set $i \leftarrow 0$.
        \begin{algorithmic}[1]
        \Repeat
             \State Solve problem (P2-2) for given $\{\alpha, \mathbf{\Phi}\}$, and update $\{\mathbf{f}\}$ as the optimal solution.
            \State Solve problem (P3) for given $\{\mathbf{f}, \mathbf{\Phi}\}$ by grid search algorithm, and update $\{\alpha\}$ as the optimal solution.
            \State Solve problem (P4-3) for given $\{\mathbf{f}, \alpha\}$, and update $\{\mathbf{\Phi}\}$ as the optimal solution.
            \State Update $i \leftarrow i+1$.
        \Until The fractional decrease of the objective value is below a threshold $\epsilon>0$.
        \end{algorithmic} 
\end{algorithm}
Based on the above procedure, we propose the AO-based Algorithm 1 to solve (P1) by alternately optimizing the RIS phase shift, the bandwidth splitting ratio and the receive beamforming.\\

\subsection{Convergence and Complexity Analyses}
For solving (P1), we propose the AO-based Algorithm 1, which guarantees that the objective function in (\ref{p1}a) monotonically decreases with each iteration \cite{Xu2024Complexity}. Consequently, the proposed algorithm ensures convergence by iterating until the fractional decrease of the objective value falls below a predefined threshold $\epsilon$, which serves as the convergence condition.

The complexity of Algorithm 1 is mainly determined by the sub-problems introduced by the AO-based method. Specifically, (P2-2), which addresses optimization of receive beamforming, is formulated as an SDR problem with complexity $\mathcal{O}(N^4K^{0.5})$ \cite{Xu2024Complexity, Luo2010Complexity,Zhang2024Complexity}. Note that the computational cost of the rank-one recovery process is negligible compared to that of the SDR procedure \cite{Luo2010Complexity}. The second sub-problem (P3), corresponds to the optimization of the bandwidth splitting ratio and is based on grid search, with complexity $\mathcal{O}(GK)$, where $G$ denotes the number of grid points evaluated, i.e., $1/\nu$. Lastly, sub-problem (P4-3), which optimizes the receive beamforming and has a similar structure to (P2-2), has a complexity of $\mathcal{O}(M^4K^{0.5}+K^{4.5})$. Therefore, the overall complexity of Algorithm 1 is given by $\mathcal{O}\big(\mathcal{I}(N^4K^{0.5}+K^{4.5}+GK+M^4K^{0.5})\big)$, where $\mathcal{I}$ represents the total number of iterations of Algorithm 1. 
\section{Numerical Results}
\label{Numerical Results}
\begin{table}[t]
    \vspace{-2pt}
    \caption{Simulation parameter setting}
    \label{parameter}
    \centering
    \vspace{-4pt}
    \begin{tabular}{cccccc} 
        \\[-1.8ex]\hline 
        \hline \\[-1.8ex] 
        \multicolumn{1}{c}{Parameter} & \multicolumn{1}{c}{Value} & \multicolumn{1}{c}{Parameter} & \multicolumn{1}{c}{Value}\\
        \hline \\[-1.8ex] 
        {$N$} & $8$   &{$f$} &  $28\times10^9\,$Hz \cite{NF_ISAC}\\
        {$K$} & $3$   &{$D$} &  $0.5\,$m \cite{NF_ISAC}\\
        {$p_\text{s}$} & $15\,$dBm  \cite{Semi_ISAC} &{$p_k$} &  $15\,$dBm \cite{Liu2023tx_power} \\
        {$B$} & $50\,$MHz    &{$\rho_\text{ISAC}$} &  $10^{-10}$\\
        {$\rho_\text{SO}$} & $10^{-9}$ \cite{psd}   &{$\sigma^2$} &  $-75$ dBm\\
        {$\sigma_{\tau}^2$} & $1.2\times 10^{-18}$   & $\beta_\text{s}$ &  $2 \times 10^{-5}$ \cite{NF_ISAC}\\
        {$B_O$} & $5 \times 10^7$   &{$T$} &  $0.1\,\mu$s\\
        {$R_\text{s}^{th}$} & $5$ Mbps\\
         
        \\[-1.8ex]\hline  
        \hline \\[-1.8ex] 
    \end{tabular}
    \vspace{-10pt}
\end{table} 

In this section, we present the numerical results to validate the performance superiority of the proposed algorithm compared to the following reference schemes.
\begin{itemize}
    \item \textit{Proposed Hybrid-Semi-ISAC}: This scheme represents our proposed algorithm, which jointly optimizes the receive beamforming vector, bandwidth-splitting ratio and RIS phase shifts. For the importance of bandwidth splitting, we consider a special case of $\alpha=0.5$, where the equal bandwidths of SO and ISAC are set.
    \item \textit{Random RIS}: The bandwidth splitting ratio and the receive beamforming are optimized by the proposed algorithm, while adopting the random phase shift matrix.  
\item \textit{Full-ISAC}: The RIS phase shift and the receive beamforming are optimized by the proposed algorithm, while exploiting only the ISAC band with $\alpha=0$ \cite{noma_ris_isac_1}. 
\item \textit{Full-SO}: The RIS phase shift and the receive beamforming are optimized by the proposed algorithm, while exploiting only the SO band with $\alpha=1$. Note that the \textit{Full-SO} represents a lower bound on the sensing performance achieved by the proposed system by ignoring data rate constraints (\ref{p1}c) and (\ref{p1}d) in (P1). \end{itemize} 

For the simulations, we adopt the standard configuration settings of 6G ISAC systems based on the 3GPP \cite{3gpp, Zhang2025TCOM_3GPP} specifications, which are applicable to both indoor and outdoor deployment scenarios \cite{IIoT_noma, Yang2024Vehicle_TWC}. The remaining parameters, e.g., the transmit power, aperture of the ULA antenna and center frequency, related to NF and FF ISAC systems are referred to existing studies \cite{Semi_ISAC, NF_ISAC, Liu2023tx_power}, as summarized in Table \ref{parameter}. To evaluate the sensing performance in a realistic RIS-assisted ISAC environment, we convert the estimation error of the time-delay estimator into centimeter-level estimation error and use it as a performance metric.

\begin{figure}[t]
    \centering
    \includegraphics[width=1\linewidth]{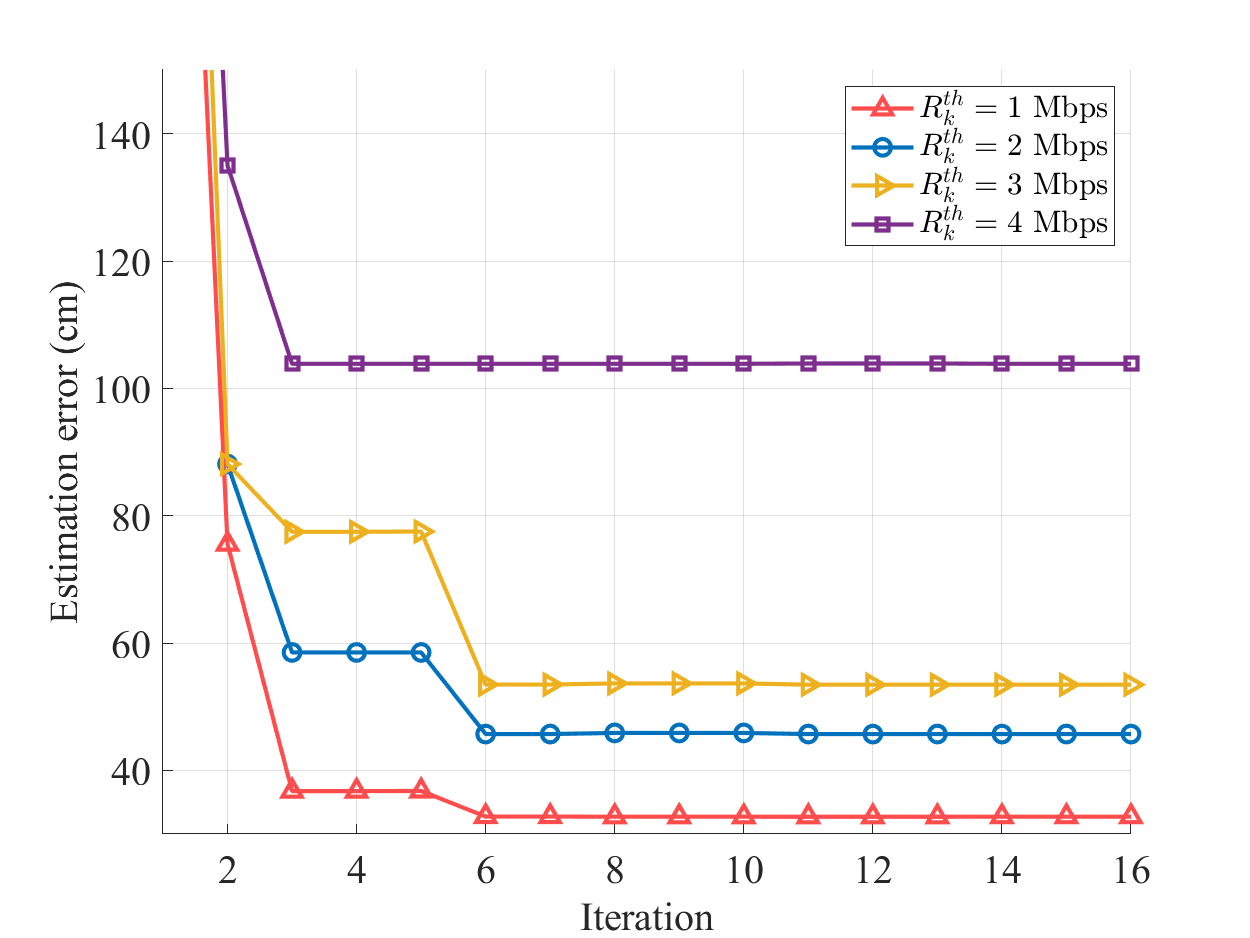}
    \caption{Convergence of proposed method according to different CU's rate threshold $R_k^{th}$ ($\mathbf{q}_\text{RIS}=(25,0,10)$, $K=3$, $N=8$, $M=32$).}
    \label{converge_result}
\end{figure}

\subsection{Convergence of Proposed Algorithm}
Fig. \ref{converge_result} illustrates the convergence behavior of the proposed method under a fair comparison setting, where the locations of the AP and IU are set to $\mathbf{q}_\textrm{AP}=(0,0,5)$, $\mathbf{q}_\textrm{s}=(20,10,0)$, respectively, while the CUs are uniform randomly distributed at a distance of $80$ m from the AP, and the RIS is deployed at $\mathbf{q}_\text{RIS}=(25,0,10)$. The AP is equipped with $N=8$ antennas along with an RIS comprising $M=32$ elements. Each case in the \textit{proposed Hybrid-Semi-ISAC} method is configured by varying the minimum rate threshold of the CUs. As shown in Fig. \ref{converge_result}, all cases converge within $16$ iterations. It is observed that a lower minimum rate threshold leads to a higher proportion of the SO band usage, thereby achieving a lower estimation error. For instance, when $R_k^{th}=4$ Mbps, the optimized bandwidth splitting ratio is $\alpha=0.41$, while for $R_k^{th}=1$ Mbps, the optimal $\alpha$ is increased to $0.94$. With optimized variables, the performance gain is attributed to the increased bandwidth allocated for the IU sensing in the SO band, which effectively enhances the estimation accuracy.


\subsection{Impact of RIS Element and CU Configuration}
\begin{figure}[t]
    \centering
    \includegraphics[width=1\linewidth]{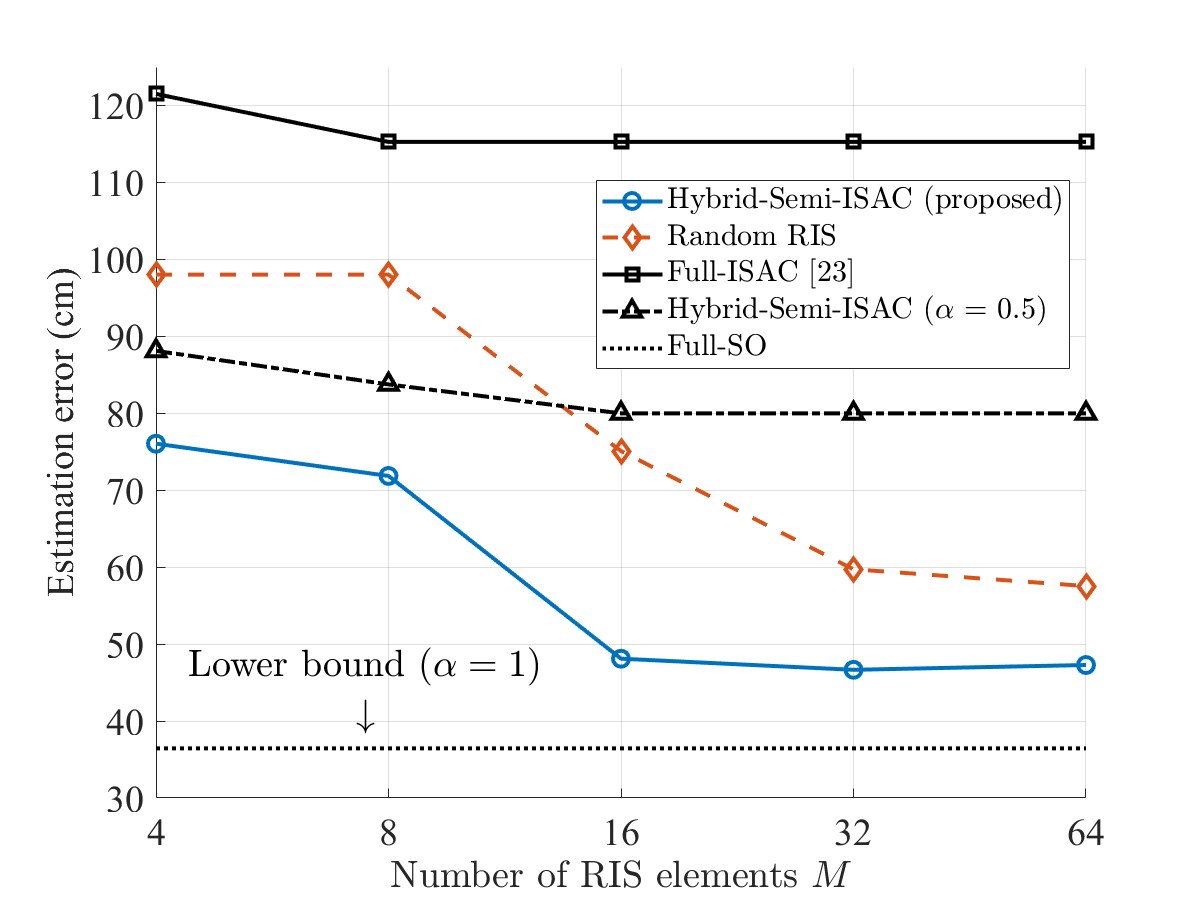}
    \caption{Estimation error performance versus the number of RIS elements $M$ ($\mathbf{q}_\text{RIS}=(25,0,10)$, $K=3$, $N=8$, $R_k^{th}=2$ Mbps).}
    \label{RIS_result}
\end{figure}
Fig. \ref{RIS_result} demonstrates the estimation error as the number of RIS elements. The \textit{Full-SO} scheme serves as a lower bound by fully allocating the SO band to minimize the estimation error of the IU. As shown in Fig. \ref{RIS_result}, both \textit{Full-ISAC} and \textit{Hybrid-Semi-ISAC with fixed $\alpha=0.5$} exhibit convergence behavior as the number of RIS elements increases. This is because, with a fixed bandwidth splitting ratio $\alpha$, it becomes challenging to achieve significant CRB performance gains solely by increasing the number of RIS elements. In contrast, the \textit{Random RIS} and the \textit{proposed Hybrid-Semi-ISAC} exhibit consistent performance improvement across the entire range of RIS elements. This observation highlights that the optimization of the bandwidth splitting ratio plays a dominant role in reducing the estimation error. Among all schemes, the proposed method achieves the best performance in all configurations of RIS elements by jointly optimizing the RIS phase shift, the receive beamforming vector and the bandwidth splitting ratio.


\begin{figure}[t]
    \centering
    \includegraphics[width=1\linewidth]{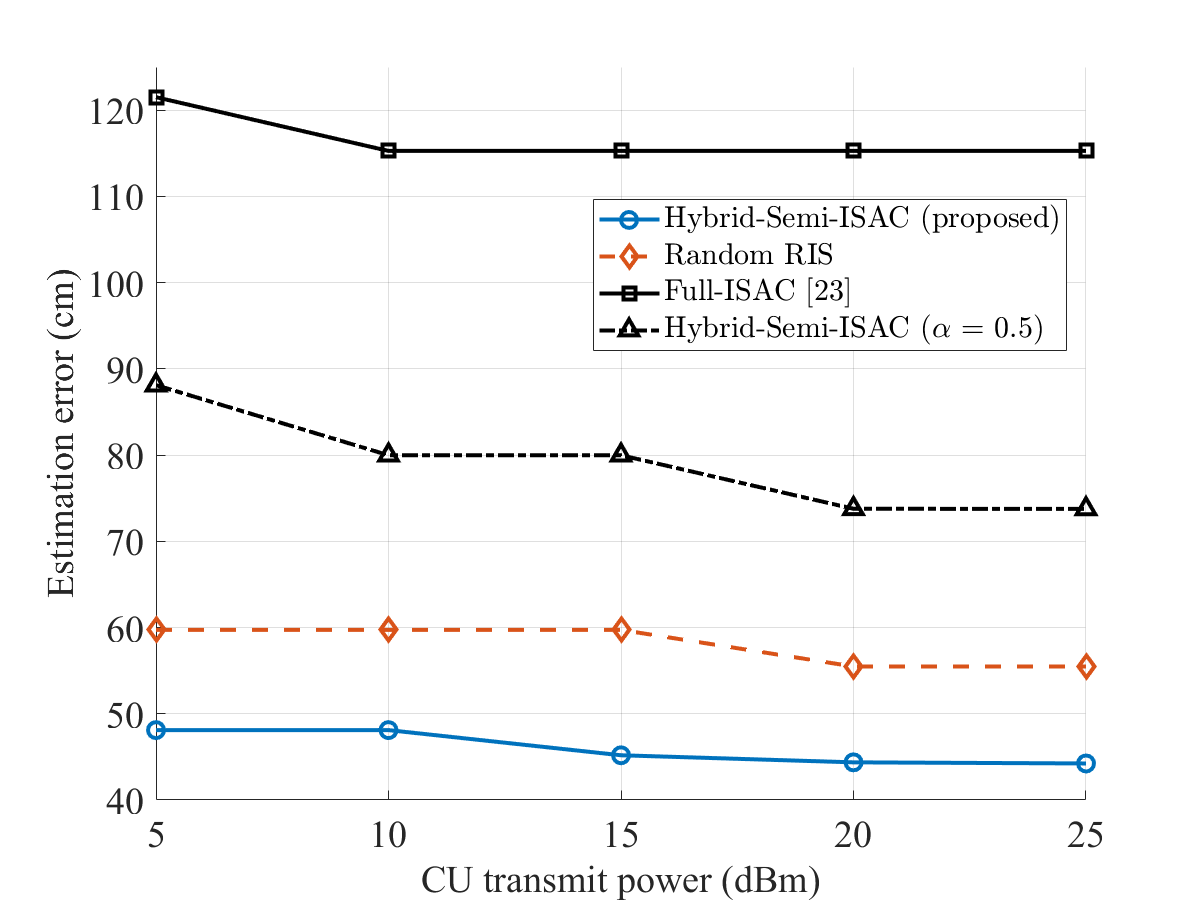}
    \caption{Estimation error performance versus CU's transmit power $p_k$ ($\mathbf{q}_\text{RIS}=(25,0,10)$, $K=3$, $N=8$, $R_k^{th}=2$ Mbps).}
    \label{p_k_result}
\end{figure}

Fig. \ref{p_k_result} demonstrates the impact of increasing the CUs' transmit power on the estimation error. All schemes exhibit improved performance as the transmit power increases. Notably, most schemes converge around near $ p_k=20$ dBm, implying that the transmit power level is sufficient to saturate the system performance under the given configuration.

\begin{figure}[t]
    \centering
    \includegraphics[width=1\linewidth]{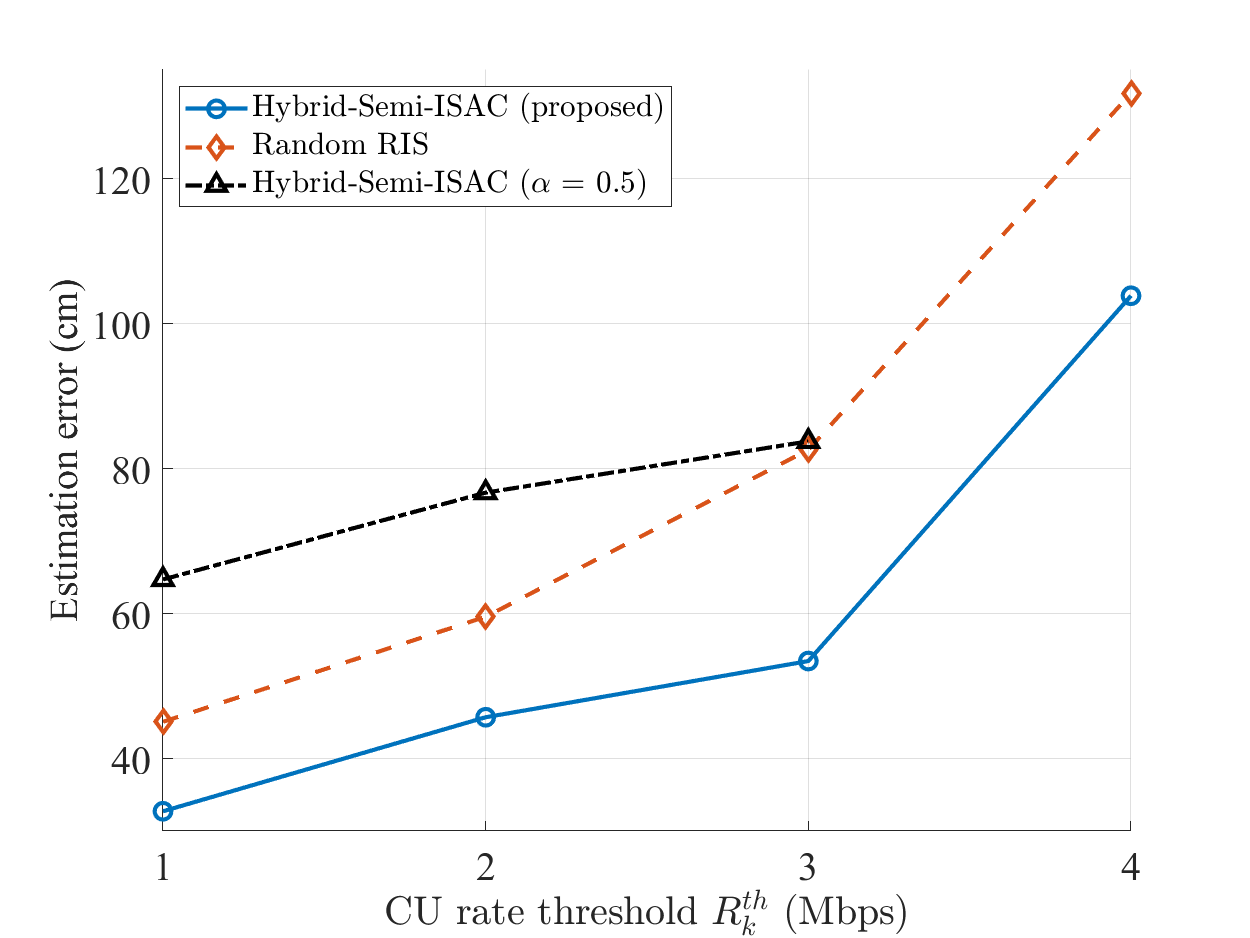}
    \caption{Estimation error performance versus CU's rate threshold $R_k^{th}$ ($\mathbf{q}_\text{RIS}=(25,0,10)$, $K=3$, $N=8$, $M=32$).}
    \label{R_k_result}
\end{figure}

Fig. \ref{R_k_result} presents the variation of the estimation error as the minimum rate requirement of the CUs increases. The proposed scheme consistently outperforms all benchmark methods across the entire threshold range. As the rate threshold increases, a larger portion of resources must be allocated to the ISAC band to satisfy the rate requirement in (\ref{p1}d), which consequently diminishes resources available for sensing, leading to higher overall estimation error. Notably, at a threshold of approximately $4$ Mbps, the \textit{Hybrid-Semi-ISAC with fixed $\alpha=0.5$} scheme fails to satisfy the requirement, highlighting the limitations in fixing the bandwidth splitting ratio to $\alpha=0.5$. In contrast, the \textit{proposed Hybrid-Semi-ISAC} method adaptively optimizes the bandwidth splitting ratio to more strictly satisfy rate constraints while reducing estimation errors.

\subsection{Impact of RIS Horizontal Position on Near/Far Field Transition and NOMA}
\begin{figure}[t]
    \centering
    \includegraphics[width=1\linewidth]{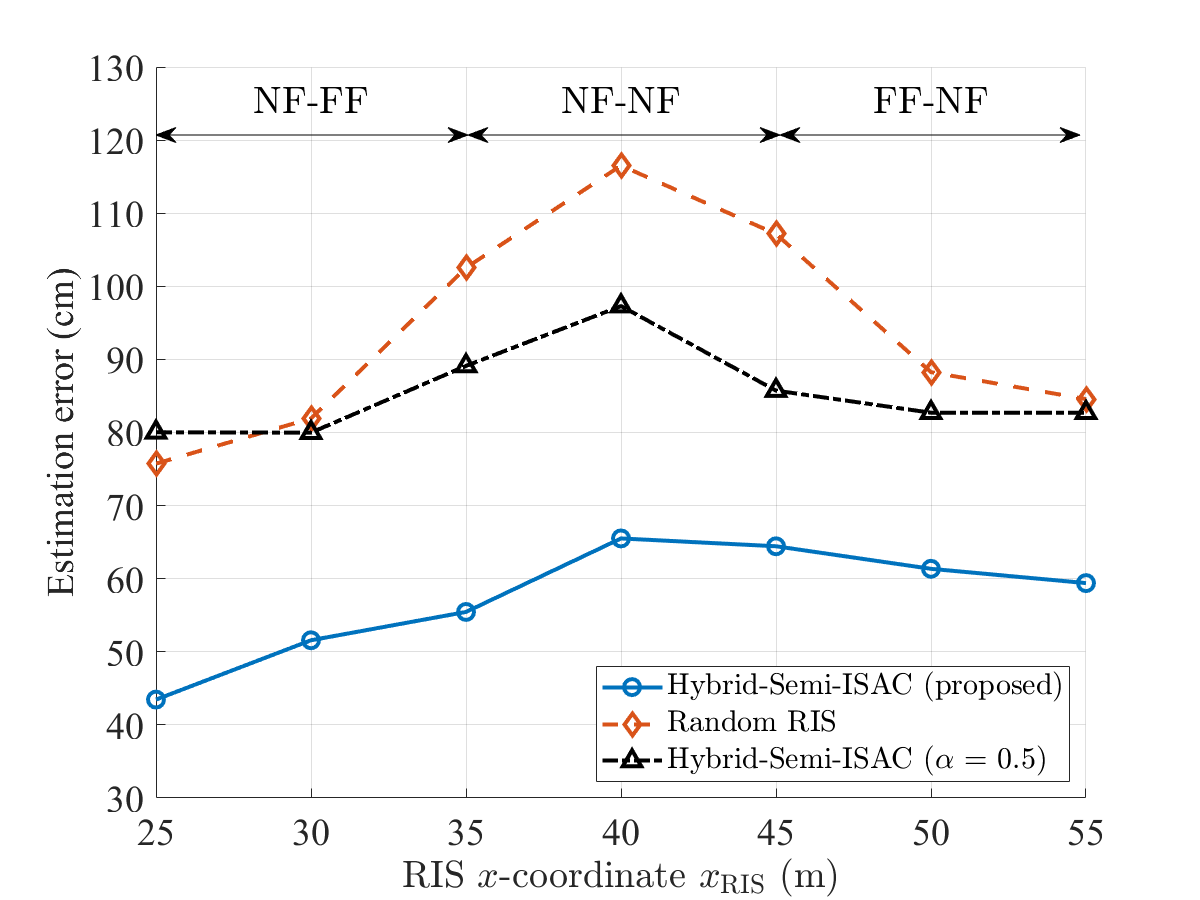}
    \caption{Estimation error performance versus RIS's $x$-coordinate $x_\text{RIS}$ ($K=3$, $N=8$, $M=32$, $R_k^{th}=2$ Mbps).}
    \label{NOMA_result}
\end{figure}

To investigate the impact of NOMA decoding performance and the transition between NF and FF channel conditions, Fig. \ref{NOMA_result} demonstrates the estimation error as a function of the RIS’s $x$-coordinate. In the simulation set-up, the RIS horizontal position is varied such that the AP–RIS and RIS–CUs links undergo transitions between NF and FF propagation regimes. Specifically, for $20\text{m}\leq x\leq30\text{m}$
, the AP–RIS link is in the NF, while the RIS–CUs link is in the FF; in the range $35\text{m}\leq x\leq45\text{m}$, both links are in the NF; and for $50\text{m}\leq x\leq55\text{m}$, the AP–RIS link becomes FF, while the RIS–CU link remains NF. It is observed that the worst sensing estimation performance occurs around $x=40$m, where severe multiplicative fading through the RIS substantially degrades the system performance.
The \textit{Random RIS} scheme shows considerable performance fluctuations and degradations due to the lack of optimized RIS beamforming. The \textit{proposed Hybrid-Semi-ISAC} method outperforms all benchmarks across the entire RIS deployment range by adaptively optimizing the RIS phase-shifts and receive beamforming and adjusting the bandwidth splitting ratio. Specifically, the proposed method shows the largest performance gap in the range $25\text{m}\leq x\leq30\text{m}$ where the AP and RIS are closely located. In the fully NF region, e.g., $35\text{m}\leq x\leq45\text{m}$, the accurate beam alignment becomes critical, and lacking phase-shift optimization in \textit{Random RIS} hence causes larger performance swings. When the RIS approaches the CUs in the $50\text{m}\leq x\leq55\text{m}$ range, system performance improves due to more favorable proximity. This performance variation ultimately reflects the sensitivity of NOMA decoding and NF and FF transition according to the RIS position, as the rate constraints imposed by CUs directly affect the residual interference.

\section{Conclusion}
\label{Conclusion}
In this paper, we have investigated an RIS-assisted ISAC framework tailored for generalized hybrid NF-FF  scenarios. To address the challenges of heterogeneous network environments, we introduce a resource allocation strategy that jointly utilizes both the ISAC and SO bands. Furthermore, a NOMA-based uplink communication protocol is employed to accommodate multiple types of ISAC and data communication users, enabling efficient sequential decoding of superimposed communication and sensing signals. To enhance overall system performance, we propose a joint design that minimizes the sensing accuracy in terms of CRB of time-delay estimation by optimizing the RIS phase shifts, bandwidth-splitting ratio and receive beamforming vector under user-specific data rate constraints. The resulting non-convex problem has been efficiently solved using an AO framework integrated with a SDR technique. Simulation results have verified that the proposed method significantly outperforms benchmark schemes relying solely on either the ISAC or SO band, equal-bandwidth allocation or random RIS configurations. In addition, we have proposed an adaptive bandwidth and beamforming management approach, providing robust sensing and communication performance under diverse NOMA and hybrid NF and FF channel conditions.

Future research can be extended to more practical and large-scale wireless environments that encompass the massive ISAC and communication user scenarios. To this end, incorporating learning-based optimization and model-driven machine learning techniques can be a promising direction to further enhance sensing accuracy and system adaptability in the presence of massive user deployments.

\appendix
\subsection{Derivation of FIM}
By applying the Parseval’s theorem for the transition from the time domain to the frequency domain in (\ref{crb_derive}), the time-domain signals $x_\text{SO}(t)$ and $x_\text{ISAC}(t)$ are transformed into their frequency-domain representations, denoted as $X_\text{SO}(f)$ and $X_\text{ISAC}(f)$, respectively. Therefore, the FIM with respect to the unknown parameter $\tau$ can be derived as 
\begin{equation}
\begin{aligned}
     &J(\alpha,\mathbf{f})\hspace{-3pt}=\hspace{-3pt}\frac{2\lVert a \rVert^2}{\sigma^2}\big[\int_{B_O-\frac{B}{2}}^{B_O-\frac{B}{2}+\alpha B}\hspace{-25pt}p_\text{SO}\langle X_\text{SO}(f)X_\text{SO}^*(f)\rangle(2\pi f)^2df\\&+\int_{B_O-\frac{B}{2}+\alpha B}^{B_O+\frac{B}{2}}p_\text{ISAC}\langle X_\text{ISAC}(f)X_\text{ISAC}^*(f)\rangle(2\pi f)^2df\bigg]\\
     &=\dfrac{8\pi^2 N^2 \beta_{\textrm{s}}^2 | \mathbf{f}^H\mathbf{a}_\text{AP,s} |^2 T B}{3\sigma^2}\times\\&\bigg[\alpha\rho_\textrm{SO}\bigg(\big(B_O-\frac{B}{2}+\alpha B\big)^3-\big(B_O-\frac{B}{2}\big)^3\bigg)\\&-(1-\alpha)\rho_\textrm{ISAC}\bigg(\big(B_O-\frac{B}{2}+\alpha B\big)^3-\big(B_O+\frac{B}{2}\big)^3\bigg)\bigg],
\end{aligned}
\end{equation}
where $B_O$ is a free parameter and is determined based on the reduced FIM as presented in \cite{Bliss}.

\subsection{Proof of Lemma 2}
Let the FIM function in (\ref{FIM_eq}) be redefined as
\begin{equation}
\begin{split}
     J(\alpha,\mathbf{f})&=A\bigg[\alpha\rho_\textrm{SO}\bigg(\big(B_O-\frac{B}{2}+\alpha B\big)^3-\big(B_O-\frac{B}{2}\big)^3\bigg)-\\&(1-\alpha)\rho_\textrm{ISAC}\bigg(\big(B_O-\frac{B}{2}+\alpha B\big)^3-\big(B_O+\frac{B}{2}\big)^3\bigg)\bigg],
\end{split}
\end{equation}
where $A\triangleq {8\pi^2 N^2 \beta_{\textrm{s}}^2 | \mathbf{f}^H\mathbf{a}_\text{AP,s} |^2 T B}/{3\sigma^2}$. To analyze the convexity of $J(\alpha)$, we compute the second derivative with respect to $\alpha$, which yields
\begin{equation}
\begin{aligned}
    &\frac{d^2J(\alpha)}{d\alpha^2}=A\bigg(4B \rho_\text{SO}(B_O-\frac{B}{2}+\alpha B)^2\\&+6\alpha B\rho_\text{SO}(B_O-\frac{B}{2}+\alpha B)+4B \rho_\text{ISAC}(B_O-\frac{B}{2}+\alpha B)^2\\&\quad\quad\quad\quad\quad\quad\quad+6(1-\alpha)B\rho_\text{ISAC}(B_O-\frac{B}{2}+\alpha B)\bigg).
\end{aligned}
\end{equation}
Since all coefficients $A,B,\rho_\textrm{SO}, \rho_\textrm{ISAC}$ and $\alpha$ are positive, and the squared terms involving $(B_0-{B}/{2}+B_0)$ dominate the linear components, the second derivative is strictly positive for $\alpha \in [0,1]$. Therefore, $J(\alpha)$ is convex with respect to $\alpha$ for $0 \leq \alpha \leq 1$.

\bibliographystyle{IEEEtran}
\bibliography{ref}

\end{document}